\documentclass[twocolumn]{aastex631}
\usepackage{array}
\usepackage{float}
\usepackage{xcolor}
\usepackage{multirow}

\begin{document}

\title{Modeling High Mass X-ray Binaries to Double Neutron Stars through Common Envelope Evolution}

\author[0009-0009-4482-6350]{Yu-Dong Nie}
\affiliation{Department of Astronomy, Nanjing University, Nanjing 210023, People's Republic of China}
\affiliation{Key Laboratory of Modern Astronomy and Astrophysics, Nanjing University, Ministry of Education, Nanjing 210023, People's Republic of China}

\author[0000-0003-2506-6906]{Yong Shao}
\email{shaoyong@nju.edu.cn}
\affiliation{Department of Astronomy, Nanjing University, Nanjing 210023, People's Republic of China}
\affiliation{Key Laboratory of Modern Astronomy and Astrophysics, Nanjing University, Ministry of Education, Nanjing 210023, People's Republic of China}

\author[0000-0003-3862-0726]{Jian-Guo He}
\affiliation{Department of Astronomy, Nanjing University, Nanjing 210023, People's Republic of China}
\affiliation{Key Laboratory of Modern Astronomy and Astrophysics, Nanjing University, Ministry of Education, Nanjing 210023, People's Republic of China}

\author[0009-0001-4454-8428]{Ze-Lin Wei}
\affiliation{Department of Astronomy, Nanjing University, Nanjing 210023, People's Republic of China}
\affiliation{Key Laboratory of Modern Astronomy and Astrophysics, Nanjing University, Ministry of Education, Nanjing 210023, People's Republic of China}

\author[0000-0002-3614-1070]{Xiao-Jie Xu}
\affiliation{Department of Astronomy, Nanjing University, Nanjing 210023, People's Republic of China}
\affiliation{Key Laboratory of Modern Astronomy and Astrophysics, Nanjing University, Ministry of Education, Nanjing 210023, People's Republic of China}

\author[0000-0002-0584-8145]{Xiang-Dong Li}
\affiliation{Department of Astronomy, Nanjing University, Nanjing 210023, People's Republic of China}
\affiliation{Key Laboratory of Modern Astronomy and Astrophysics, Nanjing University, Ministry of Education, Nanjing 210023, People's Republic of China}




\begin{abstract}

We present detailed evolutionary simulations of wide binary systems with high-mass ($8-20\,M_{\odot}$) donor stars and a $1.4\,M_{\odot}$ neutron star. Mass transfer in such binaries is dynamically unstable and common envelope (CE) evolution is followed. We use a recently developed prescription to deal with CE evolution and consider various CE ejection efficiencies varying in the range of $0.1-3.0$. We focus on the evolutionary consequences of the binaries survived CE evolution. We demonstrate that it is possible for the binaries to enter a CE decoupling phase (CEDP) when the donor stars are partially stripped leaving a hydrogen envelope of $\lesssim1.0-4.0\,M_\odot$ after CE evolution. This phase is expected to last $\sim 10^4-10^5\,\rm yr$, during which mass transfer occurs stably via Roche lobe overflow with super-Eddington rates. Identification of some X-ray binaries in a CEDP is important for the understanding of the physics of CE evolution itself, the origin of ultraluminous X-ray sources, and the recycling process of accreting pulsars. Also, we discuss the formation of double neutron stars and the occurrence of ultra-stripped supernovae according to the results from our simulations. On the whole, the properties of post-CE binaries are sensitive to the options of CE ejection efficiencies. 

\end{abstract}


\section{Introduction} \label{sec:introduction}

Since the discovery of PSR B1913+16 \citep{Hulse1975} about fifty years ago, there are more than 20 double neutron stars (DNSs) detected in the Milky Way \citep[see][for a recent compilation]{Deng2024}. The orbital periods of these DNSs are distributed in a wide range of $\sim 0.1-45$ days and about half are close binaries with orbital periods less than $\sim 1$ day. It is believed that DNSs are descendants of high mass X-ray binaries (HMXBs) with an accreting NS and a $\gtrsim 8\,M_\odot$ donor star according to the theory of binary evolution \citep{Tauris2023}. Until now, over 100 pairs of HMXBs have been observed in our Galaxy \citep{Liu2006, Fortin2023, Neumann2023}. Electromagnetic observations of radio/X-ray pulsars \citep{Joss1984, Lorimer2008} and X-ray binaries \citep{Verbunt1993} are able to provide valuable insights into the formation of DNSs. Recently, the LIGO, Virgo, and KAGRA Scientific Collaborations have successfully detected DNS mergers with gravitational wave observations  \citep{Abbott2017,Abbott2023}. It is anticipated that further observations will significantly increase the number of DNS mergers as gravitational wave sources \citep{Abbott2018, Abbott2023}. Also, DNS mergers are probably detected through electromagnetic waves \citep[][and references therein]{Metzger2017,Abbott2017b,Abbott2017c,Margutti2021}. The profound impact of DNS samples on the field of astronomy necessitates a continued investigation into the formation channels
giving rise to such systems.

\begin{figure}[htbp]
    \includegraphics[width=0.47\textwidth]{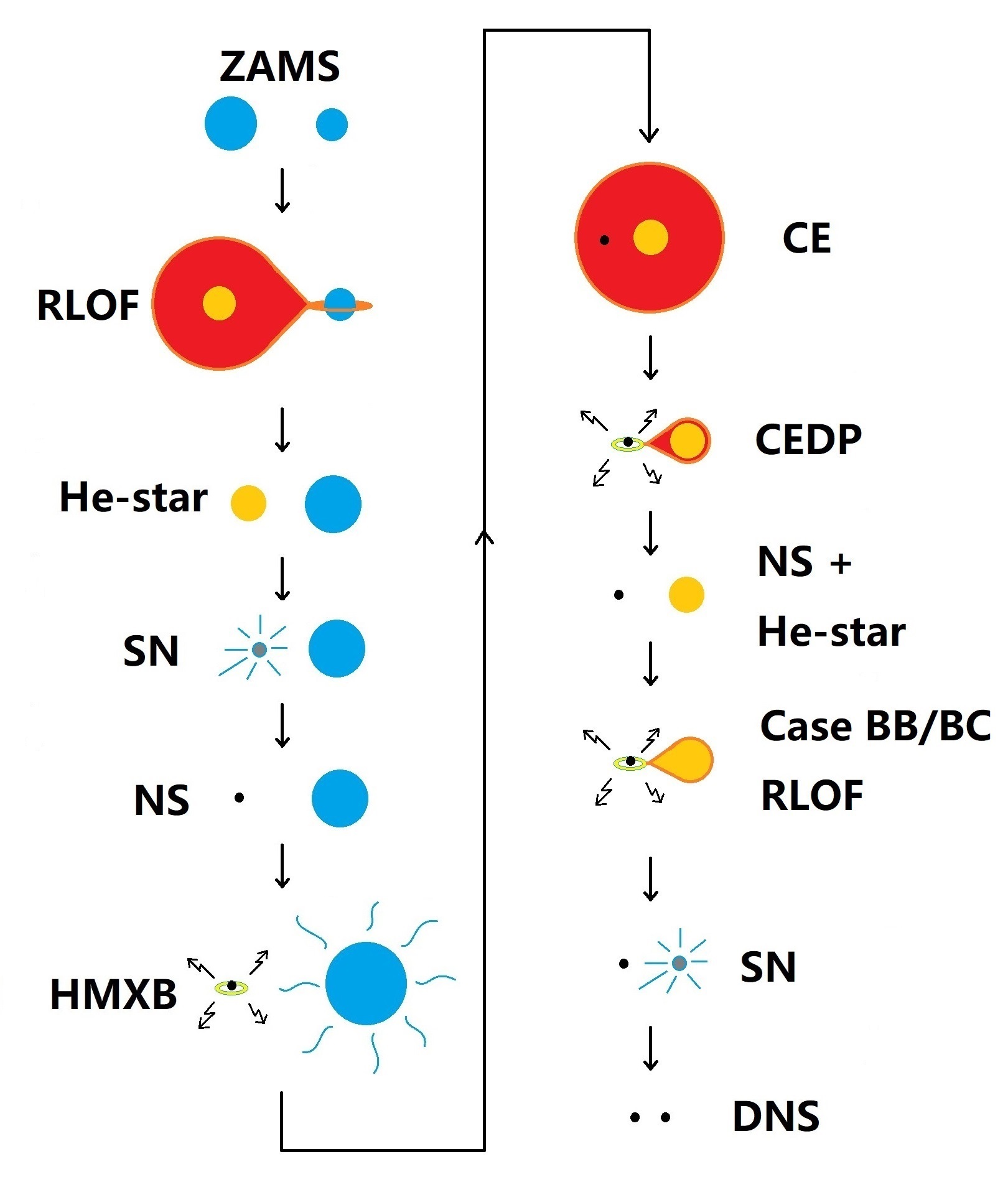}
    \centering
    \caption{Illustration of the formation of a DNS system evolved from an initial binary containing two ZAMS OB-type stars. Acronyms used in this figure - ZAMS: zero-age main-sequence; RLOF: Roche-lobe overflow; He-star: helium star; SN: supernova; NS: neutron star; HMXB: high-mass X-ray binary; CE: common envelope; CEDP: common envelope decoupling phase (see Sections \ref{sec:3.2} and \ref{sec:4.1} for the explanation of this new evolutionary phase).}
    \label{13}
\end{figure}

The canonical channel of forming DNSs has been established \citep[][]{Bhattacharya1991,Tauris2017}. Figure \ref{13} shows a schematic of this channel. Evolved from an initial binary containing two OB-type stars, it is required that both components
are massive enough to end their lives as NSs via supernova explosions. This binary must be close
enough to enable the occurrence of mass transfer (MT) between binary components, and the MT phase usually takes place via stable Roche-lobe overflow (RLOF). If the binary system remains bound after the first supernova 
explosion, the system is believed to appear as an HMXB \citep[e.g.,][]{Shao2014}. Before this stage, the binary may also be
observed as a radio pulsar orbiting an OB-star \citep[e.g.,][]{Johnston1992,Kaspi1994}. When RLOF happens in an HMXB, the process of MT is thought to be dynamically unstable. This is followed by a common-envelope (CE) phase, during which the first-born NS is engulfed by the envelope of the donor star \citep{Paczynski1976,VandenHeuvel1976}. The friction of NS's motion inside the envelope leads to a significant shrinkage of the binary orbit. If the system does not merge during the CE phase, it is
composed of an NS orbiting a helium star (the naked core of the donor star). Subsequently, an additional phase of Case BB/BC RLOF MT  may occur if the binary is tight enough. This phase allows for extreme stripping of the helium star, probably leading to the occurrence of an ultra-stripped supernova as the star explodes \citep{Tauris2015}. If the binary is not disrupted during the second supernova explosion, a DNS system forms. 

As illustrated above, CE evolution plays a vital role in understanding the formation of DNSs. Besides, CE evolution is believed to be responsible for forming tight binaries such as progenitors of Type Ia supernovae \citep{Webbink1984}, X-ray binaries \citep{Podsiadlowski2003,Shao2020} and gravitational wave sources \citep{Voss2003,Mandel2022}. Although the evolutionary scenario of CE has been proposed over forty years \citep{Paczynski1976}, binary models regarding CE evolution still have large uncertainties \citep[see][for a review]{Ivanova2013}. Important open issues for CE evolution include the trigger conditions related to MT stability \citep[e.g.,][]{Soberman1997,Ge2010,Ge2015,Ge2020,Pavlovskii2017,Han2020,Shao2021,Marchant2021} and the evolutionary outcomes related to the balance between orbital decay and CE ejection \citep[e.g.,][]{Webbink1984,Nelemans2005,Soker2015,Klencki2021,Hirai2022,DiStefano2023}.

Previous population synthesis works \citep[e.g.,][]{Tutukov1993,Andrews2015,Shao2018,Sgalletta2023,Deng2024} often used the $\alpha_{\rm CE}\lambda$ formula to deal with the orbital shrinkage of DNS's progenitor binaries during CE evolution. This formula is easy to use but crude, since it avoids a detailed treatment of the complex process of CE evolution itself \citep{Ivanova2013}. Until now, three-dimensional hydrodynamic simulations of the complete evolution of NS HMXBs via a CE phase are still challenging \citep[e.g.,][]{Moreno2022,Ropke2023,Vetter2024}. Using one-dimensional numerical calculations, an early study of \citet{Fragos2019} evolved the inspiral of a $1.4\,M_\odot$ NS inside the envelope of a $12\,M_\odot$ red supergiant star and self-consistently calculated the drag force between the NS and the donor's envelope during the inspiral of a CE phase. Due to numerical reasons, their simulation was terminated when a non-negligible hydrogen envelope still remains around the helium core. Based on the scheme recently developed by \citet{Marchant2021} that allows to calculate CE evolution by self-consistently determining the core-envelope boundary, \citet{Gallegos-Garcia2023} performed simulations for a grid of binary systems with a $1.4-2.0\,M_\odot$ NS and a $8-20\,M_\odot$ donor star until the formation of DNS mergers. In this paper, we also simulate
a grid of NS HMXBs but focus on the behavior of the donor stars during CE evolution and the properties of all post-CE systems if they survive. 

\section{Method} \label{sec:method}

We utilize the Modules for Experiments in Stellar Astrophysics (MESA) code \citep[version 12115,][]{Paxton2011,Paxton2013,Paxton2015,Paxton2018,Paxton2019} to conduct our simulations. Our models are computed beginning from the binary systems with a zero-age main-sequence star and an NS with initial mass $M^{\rm i}_{\rm NS}=1.4\,M_{\odot}$. In reality, the initial donor should be a rejuvenated star due to mass accretion during HMXB's progenitor evolution (see Figure \ref{13} or Figure \ref{12} in the Appendix \ref{AA}). The simulations involve a grid of initial orbital periods ranging between
$2.5\leq \log (P_{\rm orb}^{\rm i}/\rm d)\leq 3.5$\footnote{Throughout this paper, we use $\log$ to represent $\log_{\rm 10}$.} in steps of 0.01 and initial donor masses $M^{\rm i}_{\rm d}$ between 8\,$M_{\odot}$ and 20\,$M_{\odot}$ with an interval of $1\,M_{\odot}$. For initial donor stars, we set metallicity to be $Z=Z_{\rm \odot}$ where $Z_{\rm \odot}=0.0142$ \citep{Asplund2009}.
The orbital configuration is assumed to be initially circular, and the NS is treated as a point mass. Each simulation is terminated when the donor reaches core carbon depletion with central $^{12}$C abundance below $10^{-2}$.  On the one hand, it is challenging to model the subsequent evolution of the donor until iron core collapse \citep{Jiang2021}. On the other hand, the timescale between core carbon depletion and iron core collapse is negligible in terms of stellar and core mass evolution \citep[e.g.,][]{Woosley2002}. We expect that this treatment will not change our main conclusions. 

In our simulations, we deal with convection using the mixing length theory of \citet{Bohm1958} with a mixing length parameter $\alpha_{\rm MLT}=1.93$. Convection regions are determined with the Ledoux criterion \citep{Ledoux1947}. And, we include
convective core overshooting with the default overshooting parameter set in the MESA code. We take nuclear reaction rates from \citet{Cyburt2010} and \citet{Angulo1999}. We adopt the Dutch prescription to treat stellar winds, as proposed by \citet{Glebbeek2009}. This
prescription incorporates various submodels for different stellar types. Specifically, for stars with effective
temperature of $T_{\rm eff}>10^{4}$ K and surface hydrogen mass fraction of $X_{\rm H}>0.4$, we adopt the rates proposed by
\citet{Vink2001}. For stars with $T_{\rm eff}>10^{4}$~K and $X_{\rm H}<0.4$ (Wolf-Rayet stars), we employ the prescription developed by \citet{Nugis2000}. For stars with $T_{\rm eff}<10^{4}$ K, we use the rates of 
\citet{deJager1988}.

We follow the prescription of \citet{Hurley2002} to calculate wind accretion rate by the NS which is based on Bondi-Hoyle mechanism \citep{Bondi1944},
\begin{equation}
\dot M_{\rm NS}=-\frac{1}{\sqrt{1-e^{\rm 2}}}\left(\frac{GM_{\rm NS}}{v^{\rm 2}_{\rm w}}\right)^{2}\frac{\alpha_{\rm w}}{2a^{\rm 2}}\frac{1}{(1+v^{\rm 2})^{\rm \frac{3}{2}}}\dot M_{\rm w},
\end{equation}
where
\begin{equation}
v^{\rm 2}=\frac{v^{\rm 2}_{\rm orb}}{v^{\rm 2}_{\rm w}},
\end{equation}
\begin{equation}
v^{\rm 2}_{\rm orb}=\frac{G(M_{\rm d}+M_{\rm NS})}{a},
\end{equation}
\begin{equation}
v^{\rm 2}_{\rm w}=2\beta_{\rm w}\frac{GM_{\rm d}}{R_{\rm d}}.
\end{equation}
Here $G$ is the gravitational constant, $\dot{M}_{\rm w}$ is the mass loss rate of the donor star due to a stellar wind, $e$ and $a$ are the eccentricity and the separation of the binary orbit, and  $M_{\rm d}$ and $R_{\rm d}$ are the mass and the radius of the donor star, respectively. And, we set $\beta_{\rm w}=1/8$
and $\alpha_{\rm w}=3/2$ \citep{Hurley2002}.

For RLOF MT, we use the formula fitted by \citet{Eggleton1983} to calculate the Roche-lobe radius of the donor star and adopt the method proposed by \citet{Marchant2021} to model the process of MT. Mass accretion onto an NS is assumed to be limited by the Eddington rate, and the excess material escapes from the binary system carrying away the specific orbital angular momentum of the NS.

A CE phase is assumed to be triggered if the MT rate exceeds a threshold of $\dot{M}_{\rm CE} =1\,M_{\odot}\rm yr^{-1}$. Considering that the physical processes involved in CE evolution are still not
fully understood, we need to make some reasonable assumptions to numerically deal with this phase. We
employ the standard energy conservation prescription of \citet{Webbink1984} to compute the orbital evolution of a binary in a CE phase. According to this prescription,
the change in orbital energy resulting from the inspiral process tends to eject the envelope,
\begin{equation}
E_{\rm bind}=\alpha_{\rm CE}\Delta E_{\rm orb}.
\end{equation}
Here $\alpha_{\rm CE}$ is the CE ejection efficiency, with which the orbital energy is used to unbind the envelope of the donor star. In our simulations, we adopt $\alpha_{\rm CE}=0.1$, 0.3, 1.0, and 3.0 to test their effects. To calculate the change of orbital energy during inspiral, we employ
\begin{equation}
\Delta E_{\rm orb}=-\frac{GM_{\rm d,f}M_{\rm NS,f}}{2a_{\rm f}}+\frac{GM_{\rm d,i}M_{\rm NS,i}}{2a_{\rm i}},
\end{equation}
where $\rm i$ and $\rm f$ represent the initial and final stages of CE evolution, respectively. In Equation (5), $E_{\rm bind}$ refers to the binding energy of donor's envelope,
\begin{equation}
E_{\rm bind}=\int_{M_{\rm core}}^{M_{\rm d,i}} (-\frac{Gm}{r}+\alpha_{\rm th}u) dm.
\end{equation}

In our simulations, the specific internal energy of the gas at a given mass coordinate, denoted as $u$, plays a
role in determining the efficiency with which the envelope can be ejected during a CE phase. To quantify this efficiency,
a free parameter known as $\alpha_{\rm th}$ was introduced by \citet{Han1995}. Here, we
adopt $\alpha_{\rm th}=1.0$. We set $M_{\rm core}$ as the current mass of the donor star and model CE evolution at each evolutionary step   \citep[see also Section 2.2 of][]{Marchant2021}. These treatments allow the evolution of a CE phase to reach the point at which the donor star would contract inside its Roche lobe.

During a CE phase we force a mass-loss rate for the envelope, which is dependent on the donor's radius $R_{\rm d}$ and its Roche-lobe radius $R_{\rm RL}$. If $R_{\rm d} > R_{\rm RL}$, we set a high MT rate of $\dot{M}_{\rm high} = \dot{M}_{\rm CE}$ which approximately corresponding to MT on the adiabatic timescale of the envelope. If $(1-\delta)R_{\rm RL} < R_{\rm d} < R_{\rm RL}$, we follow \citet{Marchant2021} to interpolate an MT rate between $\dot{M}_{\rm high}$ and $\dot{M}_{\rm low}$. Here, the parameter $\delta$ is set to be 0.02 and $\dot{M}_{\rm low} = 10^{-5}M_{\odot}\,\rm yr^{-1}$ represents a low MT rate which is comparable to MT occurring on the nuclear timescale of the donor star. As the donor star contracts reaching the condition of  $R_{\rm d} < (1-\delta)R_{\rm RL}$, we assume the CE phase finishes. Using $X_{\rm center}$ and $Y_{\rm center}$ to represent the mass fractions of hydrogen and helium at the center of the donor star, and  X and Y to represent the mass fractions of hydrogen and helium at a specific point of the donor star, respectively, we identify the radius $R_{\rm spe}$ at the point of the region where $|X-X_{\rm center}|<0.01$ and $|Y-Y_{\rm center}|<0.01$ apply. We assume that CE evolution results in a binary merger, if the donor's core within $R_{\rm spe}$ overflows its Roche lobe or the orbital period of the binary reaches the minimum value of 5 minutes set by default during CE.

\section{Result} \label{sec:result}

\begin{figure*}[htbp]
    \includegraphics[width=0.89\textwidth]{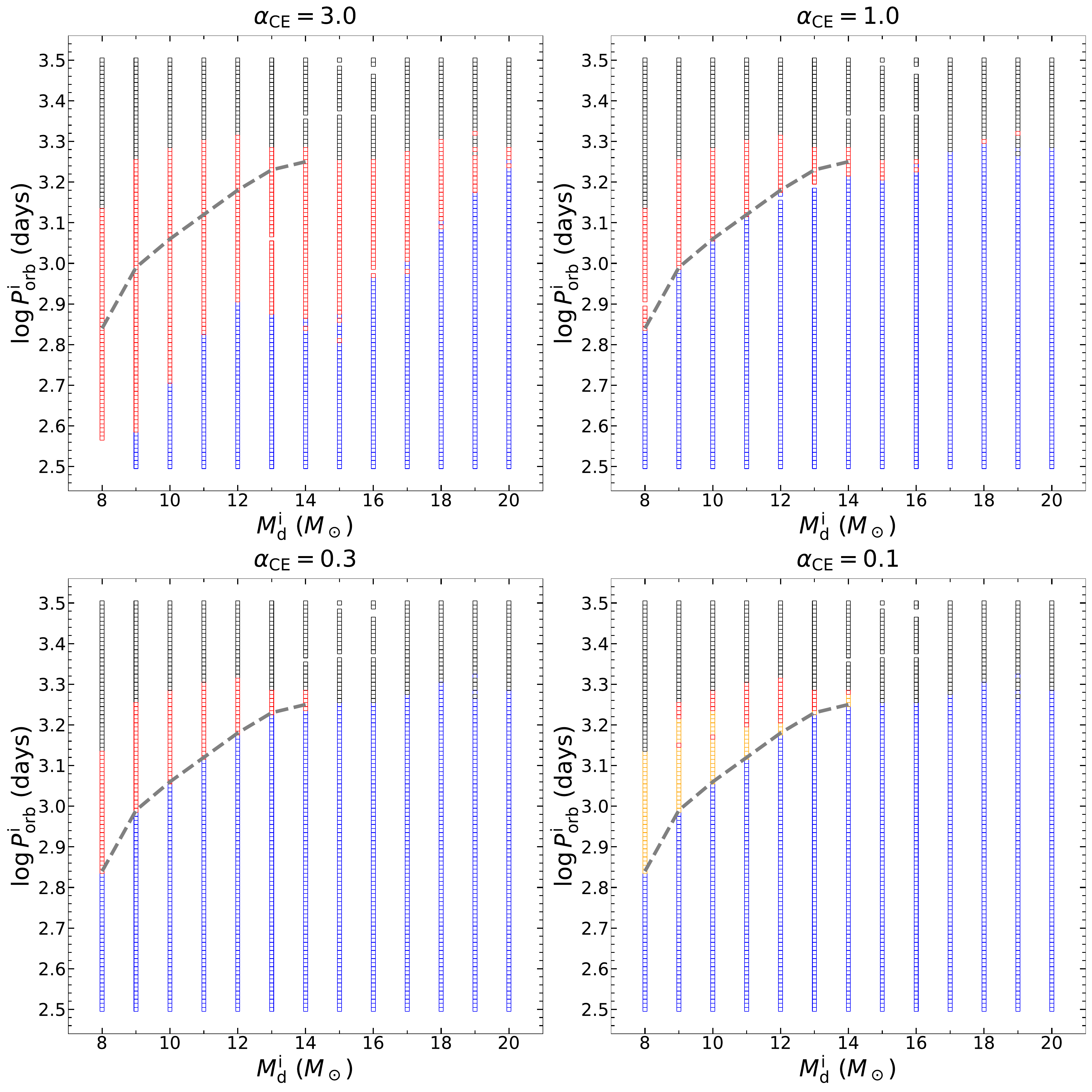}
    \centering
    \caption{Different evolutionary fates for simulations of HMXBs containing a $1.4\,M_\odot$ NS in the plane of $M^{\rm i}_{\rm d}- P^{\rm i}_{\rm orb}$. The four panels correspond to the cases with $\alpha_{\rm CE}=3.0$, 1.0, 0.3 and 0.1. In each panel, the blue, red and black squares denote the binaries being classified as CE mergers, CE survivors and noninteracting systems, respectively. Due to numerical issues, some squares do not appear to show possible evolutionary fates (especially for some wide binaries with $M_{\rm d}^{\rm i}=14-16\,M_{\rm \odot}$). The gray dashed curves show the boundaries to distinguish whether the binaries experience Case B or Case C MT, and all RLOF binaries with $M_{\rm d}^{\rm i}\geq 15\,M_{\rm \odot}$ from our simulations undergo Case B MT. In the bottom right panel, the orange squares represent CE mergers that experienced Case C RLOF MT.}
    \label{1}
\end{figure*}

\begin{figure*}[htbp]
    \includegraphics[width=0.9\textwidth]{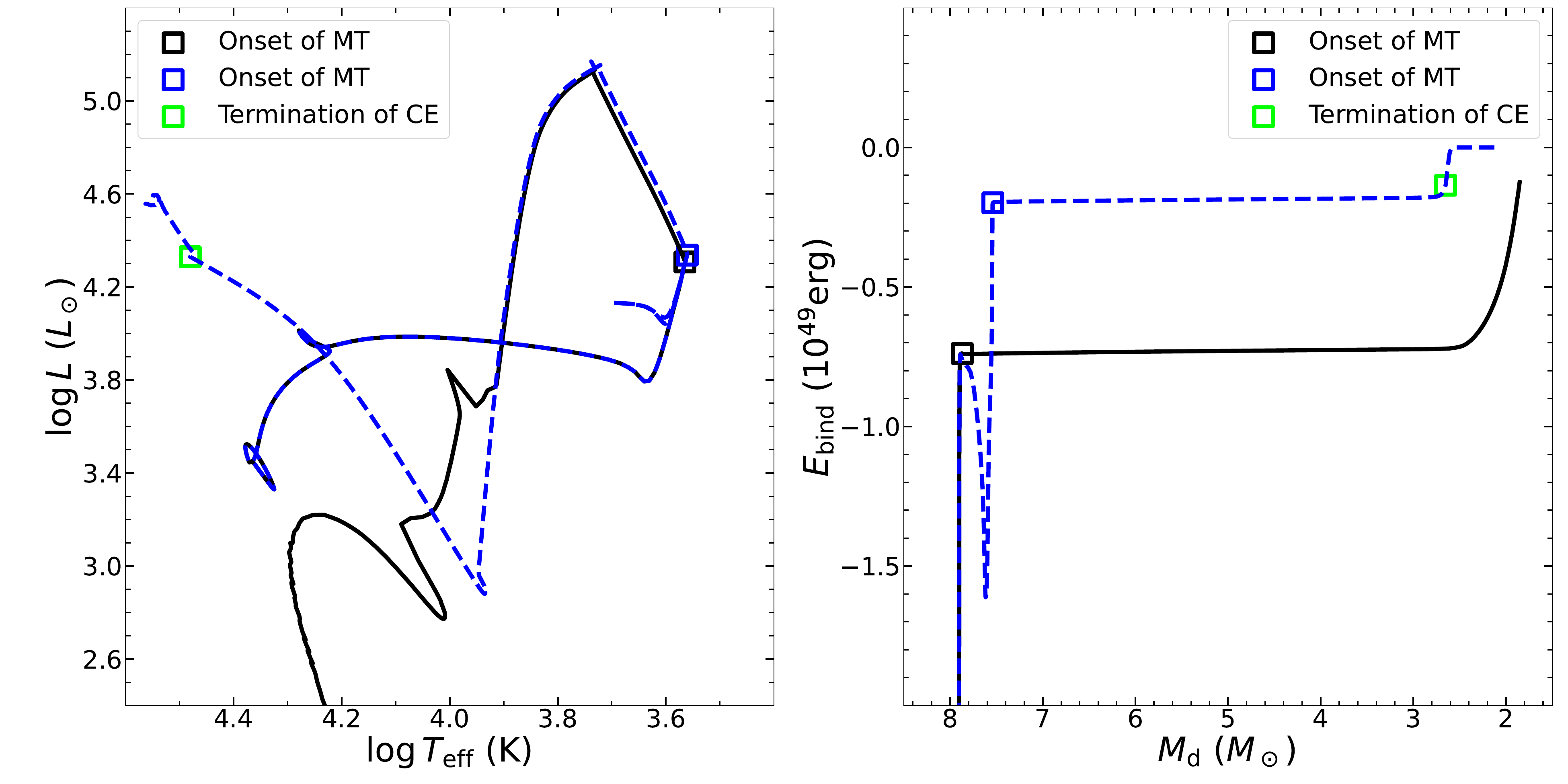}
    \centering
    \caption{Left panel: Hertzsprung-Russell diagram for the donor star of the CE merger with $M_{\rm d}^{\rm i}=8\,M_{\rm \odot}$ and $\log (P_{\rm orb}^{\rm i}/\rm d)=2.83$ (the black solid curve) or of the CE survivor with $M_{\rm d}^{\rm i}=8\,M_{\rm \odot}$ and $\log (P_{\rm orb}^{\rm i}/\rm d)=2.84$ (the blue dashed curve). Right panel: Evolution of envelope's binding energy of the donor star as a function of its mass. In each panel, three squares with different colors mark the positions at the onset of RLOF MT or at the termination of CE evolution.}
    \label{2}
\end{figure*}

\begin{figure*}[htbp]
    \includegraphics[width=0.9\textwidth]{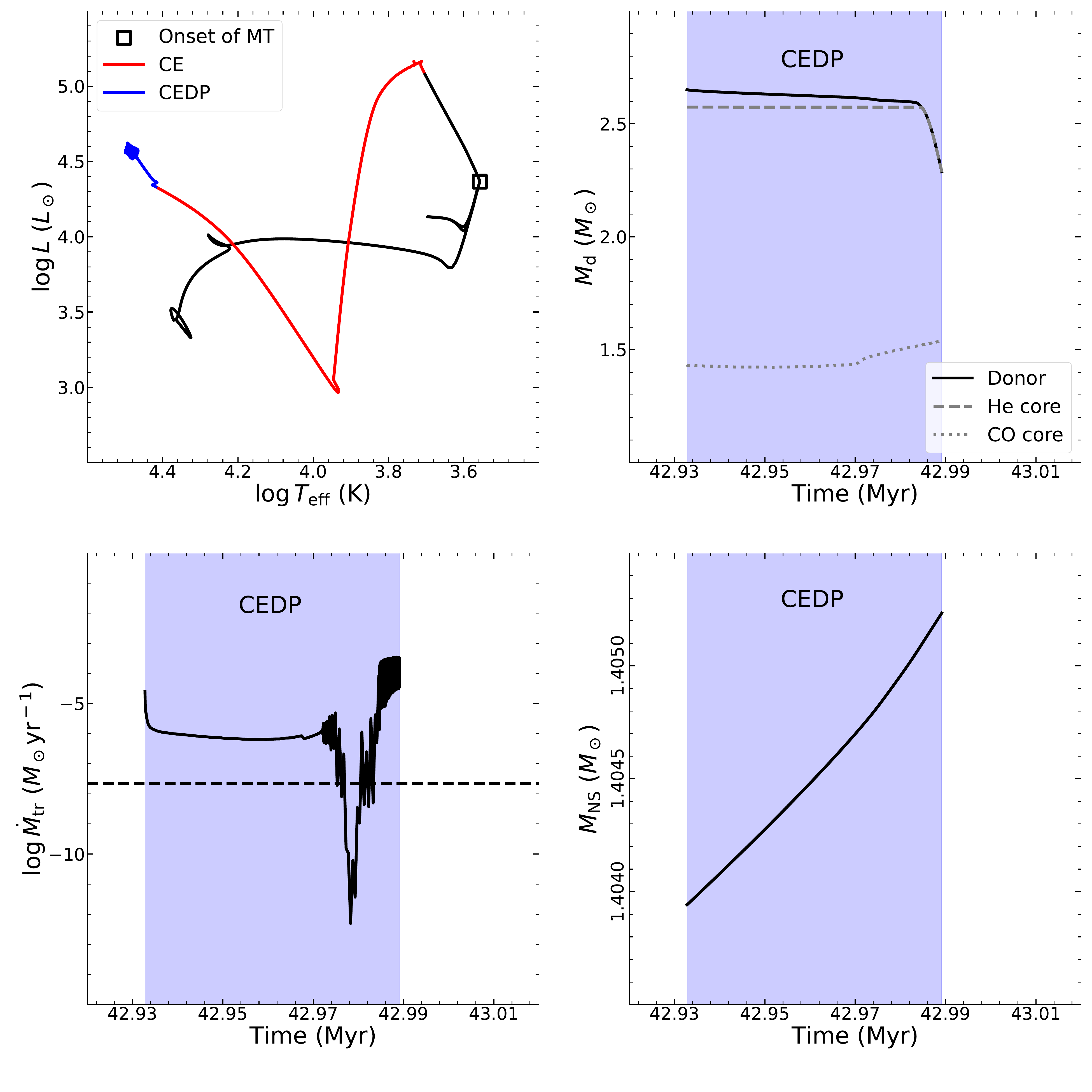}
    \centering
    \caption{Evolutionary tracks for the binary with a $1.4M_{\odot}$ NS and a $8M_{\odot}$ donor in a 741-day orbit. Here, we adopt $\alpha_{\rm CE}=1.0$. Upper left panel: Hertzsprung-Russell diagram for the donor star. We mark the position at the onset of MT with the symbol of a black square. The red and blue solid curves correspond to the binary undergoing CE and CEDP, respectively. Upper right panel: Mass of the donor star/the helium (He) core/the carbon-oxygen (CO) core as a function of time after the binary finished CE evolution. Lower left panel: Evolution of RLOF MT rate as a function of time. The black dashed line  represents the Eddington  limit. Lower right panel: Mass of the NS as a function of time.}
    \label{3}
\end{figure*}

When decreasing CE ejection efficiencies $\alpha_{\rm CE}$ from 3.0 to 0.1, Figure \ref{1} presents various evolutionary fates of the binaries we computed in the plane of $M^{\rm i}_{\rm d}$ versus $ P^{\rm i}_{\rm orb}$, which are classified as follows.

\begin{enumerate}
\item[(1)] CE mergers - the systems merge during CE evolution, as plotted with blue/orange squares. The blue and orange squares represent the CE mergers experienced Case B and Case C RLOF MT, respectively.
\item[(2)] CE survivors - the systems survive CE evolution and evolve to be close binaries with envelope-stripped donors, as plotted with red squares.
\item[(3)] Noninteracting binaries - the systems do not undergo any RLOF interaction during the whole evolution, as plotted with black squares.
\end{enumerate}
\subsection{The \texorpdfstring{$(M^{\rm i}_{\rm d},\,\log P^{\rm i}_{\rm orb})$} P Parameter Space} \label{sec:3.1}

We can see in Figure \ref{1} that CE survivors emerge in the region where $2.57\leq \log (P_{\rm orb}^{\rm i}/\rm d)\leq 3.32$ across all our adopted $\alpha_{\rm CE}$. There is a tendency that CE survivors occupy a smaller parameter space with decreasing $\alpha_{\rm CE}$. It is known that the orbital periods of observed HMXBs are always less than $\sim300$ days \citep{Liu2006}. Based on our current simulation, almost all of observed HMXBs probably evolve to be CE mergers.

For CE survivors, the maximum $M_{\rm d}^{\rm i}$ drops from $\sim 20\,M_\odot$ to $\sim 14\,M_\odot$ when decreasing $\alpha_{\rm CE}$ from 3.0 to 0.1. In the cases of $\alpha_{\rm CE}=1.0$ and $\alpha_{\rm CE}=0.3$, the lower boundaries of $\log P_{\rm orb}^{\rm i}$ remain similar shapes which generally correspond to whether the binaries experience Case B or Case C MT (RLOF MT occurs when the donor star is during the stage of shell hydrogen burning or after core helium burning). Above these boundaries, almost all CE survivors are post-Case C binaries. In the case of $\alpha_{\rm CE}=3.0$, quite a fraction of CE survivors have smaller $\log P_{\rm orb}^{\rm i}$, meaning that they are post-Case B binaries. In the case of $\alpha_{\rm CE}=0.1$, all CE survivors are post-Case C binaries having the smallest initial parameter space with $M_{\rm d}^{\rm i}\sim 9-14\,M_\odot$ and $\log (P_{\rm orb}^{\rm i}/\rm d)\sim 3.1-3.3$. In this case, part of CE mergers underwent Case C MT during the evolution (as shown by the orange squares), indicating that an NS merges with a carbon-oxygen core rather than a helium core. These mergers may lead to violent explosions such as peculiar types of gamma-ray bursts or supernovae \citep[e.g.,][and references therein]{Grichener2024}.  

In order to have a close look at the boundaries of $\log P_{\rm orb}^{\rm i}$ for distinguishing between CE survivors and CE mergers, we select two binaries with the same donor mass ($M_{\rm d}^{\rm i}=8\,M_{\odot}$) but different orbital periods ($\log (P_{\rm orb}^{\rm i}/\rm d)=2.83$ or
$\log (P_{\rm orb}^{\rm i}/\rm d)=2.84$). Under the assumption of $\alpha_{\rm CE}=1.0$, Figure \ref{2} shows the evolution of the donor star in the Hertzsprung-Russell diagram (left panel) and the envelope's binding energy $E_{\rm bind}$ of the donor star as a function of the donor mass (right panel). The binary with $\log (P_{\rm orb}^{\rm i}/\rm d)=2.83$ evolves to merge during CE evolution, while the one with $\log (P_{\rm orb}^{\rm i}/\rm d)=2.84$ is able to survive. At the onset of MT, the donor in the former is experiencing shell hydrogen burning (i.e., Case B RLOF) while the donor in the latter has formed a carbon-oxygen core (i.e., Case C RLOF). It can be seen that $|E_{\rm bind}|\sim 7\times 10^{48}\rm\,erg$ in the former is about $3-4$ times larger than that ($\sim 2\times 10^{48}\rm\,erg$) in the latter. As a consequence, the CE survivors in the cases with $\alpha_{\rm CE}=1.0$ and $\alpha_{\rm CE}=0.3$ have similar lower boundaries of $\log P_{\rm orb}^{\rm i}$. When changing $\alpha_{\rm CE}$ to 3.0 or 0.1, the corresponding lower boundaries are shown to shift significantly.

\subsection{A Case Study: \texorpdfstring{$M^{\rm i}_{\rm d}=8M_{\odot} $ and $ \log (P^{\rm i}_{\rm orb}/\rm d)=2.87$} d}
\label{sec:3.2}

Figure \ref{3} shows the evolutionary tracks of an HMXB initially containing a $1.4\,M_\odot$ NS and a $8\,M_\odot$ donor in a 741-day orbit. Here we adopt $\alpha_{\rm CE}=1.0$. The top-left panel shows the evolution of the donor star in the Hertzsprung-Russell diagram. In this section, we focus on the evolution of the binary after finishing the CE phase. The
top-right, bottom-left, and bottom-right panels respectively correspond to the post-CE evolution of the masses of the donor star and its cores, the MT rate, and the NS mass as a function of time. 

At the time of $\sim 42.93\,\rm Myr$, the donor star in the HMXB has evolved to enter the stage of asymptotic giant branch and RLOF MT starts. At this moment, a stellar wind has led to the donor star losing $\sim 0.5\,M_{\odot}$ envelope and the NS accreting $\sim 0.004\,M_{\odot}$ matter. Subsequently, MT rate rapidly increases to $1.0\,M_{\odot}\,\rm yr^{-1}$ and a CE phase is triggered. 
During this phase, $\sim 4.6\,M_{\odot}$ donor's envelope is stripped by the NS and the orbital period of the binary system drops to $\sim$ 3.9 days. 

When the donor star contracts to satisfy the condition of  $R_{\rm d} < (1-\delta)R_{\rm RL}$ and CE evolution finishes, $\sim 0.1\,M_{\odot}$ hydrogen envelope is still remained. The evolution of this binary is followed by a CE decoupling phase (CEDP), during which stable MT can last $\sim 6\times 10^4\rm\, yr$ with an averaged rate of $\sim 10^{-6}-10^{-5}M_{\odot}\,\rm yr^{-1}$. In this CEDP, such a high MT rate of the binary is able to make it an ultraluminous X-ray source \citep[ULX,][]{Fragos2019}, and $\sim 0.001M_{\odot}$ matter is accreted by the NS \citep[see also][]{Guo2024}. At the end of this phase, the donor star has lost its whole hydrogen envelope and evolved to be a $\sim 2.3\,M_\odot$ helium star with a $\sim 1.5\,M_\odot$ carbon-oxygen core. The evolution is terminated due to central carbon depletion of the donor star, the orbital period of the binary finally decreases to $\sim$ 3.3 days. 

\begin{figure*}[htbp]
    \includegraphics[width=0.9\textwidth]{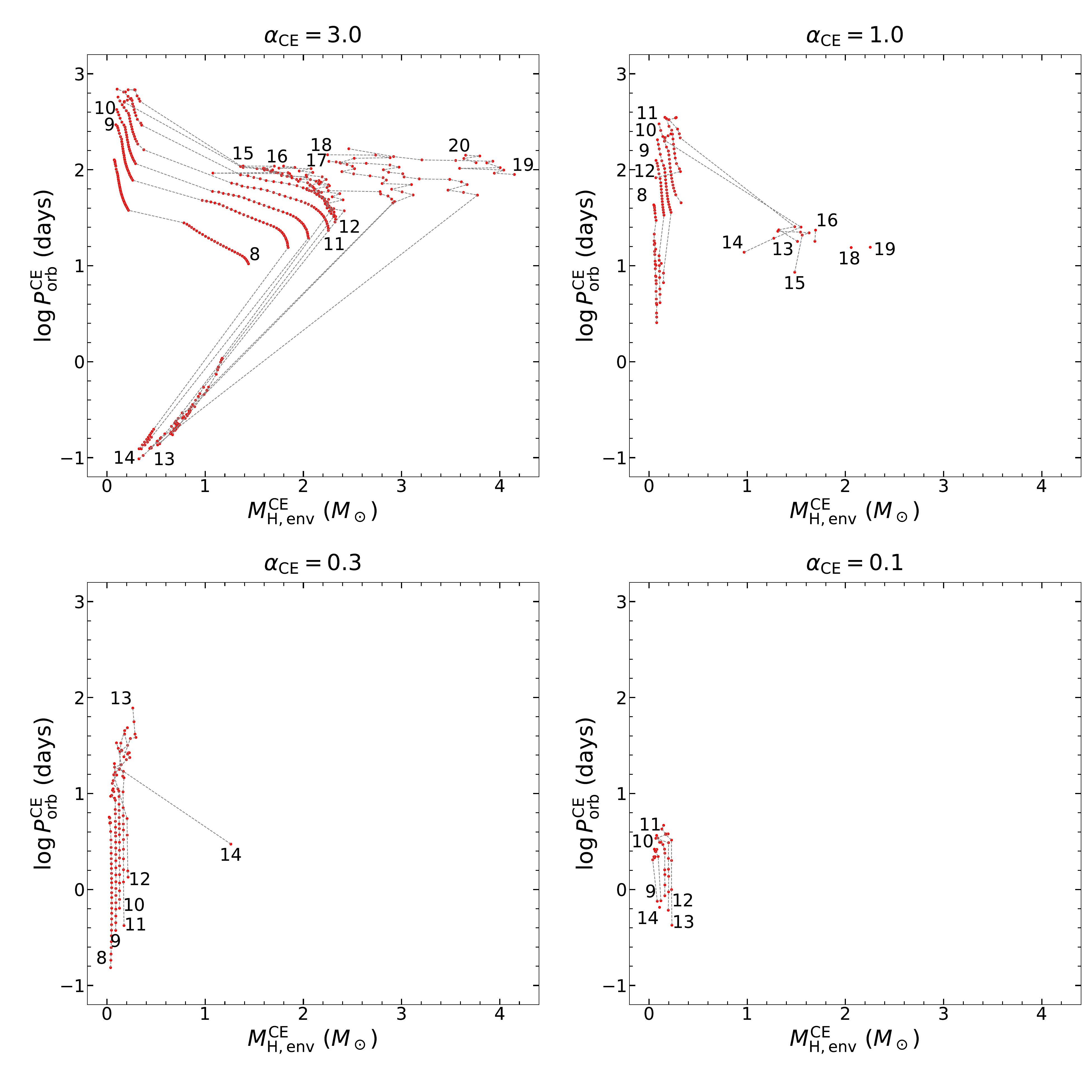}
    \centering
    \caption{Orbital period $ P^{\rm CE}_{\rm orb}$ as a function of donor's hydrogen envelope mass $M^{\rm CE}_{\rm H,env}$ at the termination of CE evolution. The four panels correspond to the cases with $\alpha_{\rm CE}=3.0$, 1.0, 0.3 and 0.1. The number next to each curve gives the initial mass of the donor star adopted in our calculations. The red dots denote all of our calculated systems that survived CE evolution.}
    \label{4}
\end{figure*}
\subsection{The \texorpdfstring{$M_{\rm H,env}^{\rm CE}-\log P_{\rm orb}^{\rm CE}$} D Diagram at the Termination of CE Evolution} \label{sec:3.3}

\begin{figure*}[htbp]
    \includegraphics[width=0.9\textwidth]{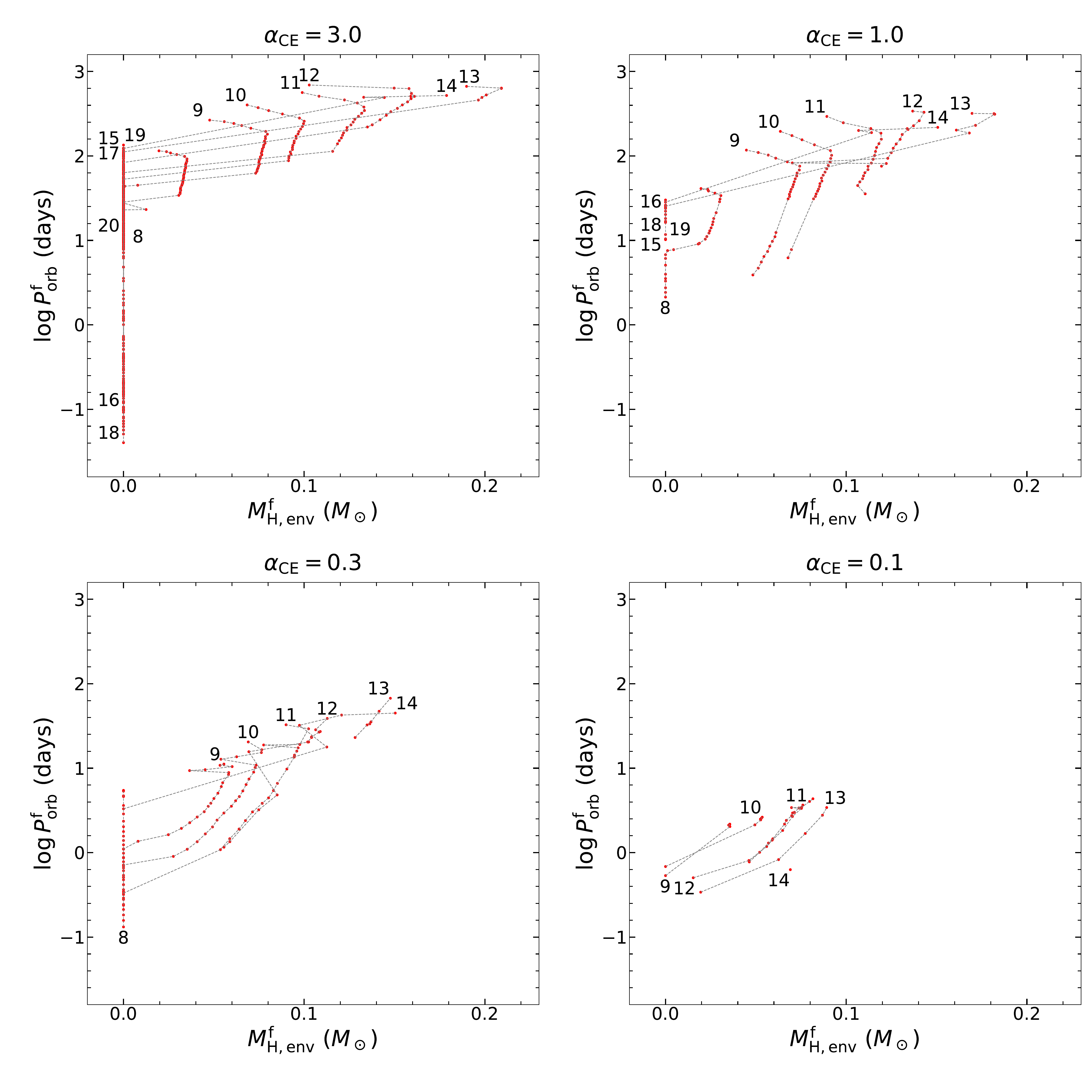}
    \centering
    \caption{Final orbital period $ P^{\rm f}_{\rm orb}$ as a function of final hydrogen envelope mass $M^{\rm f}_{\rm H,env}$ when the donor reaches central carbon depletion, by assuming four different CE ejection efficiencies with $\alpha_{\rm CE}=3.0$, 1.0, 0.3, and 0.1. The number next to each curve gives the initial mass of the donor star adopted in our calculations. The red dots denote all of our calculated systems that survived CE evolution.}
    \label{5}
\end{figure*}

\begin{figure*}[htbp]
    \includegraphics[width=0.9\textwidth]{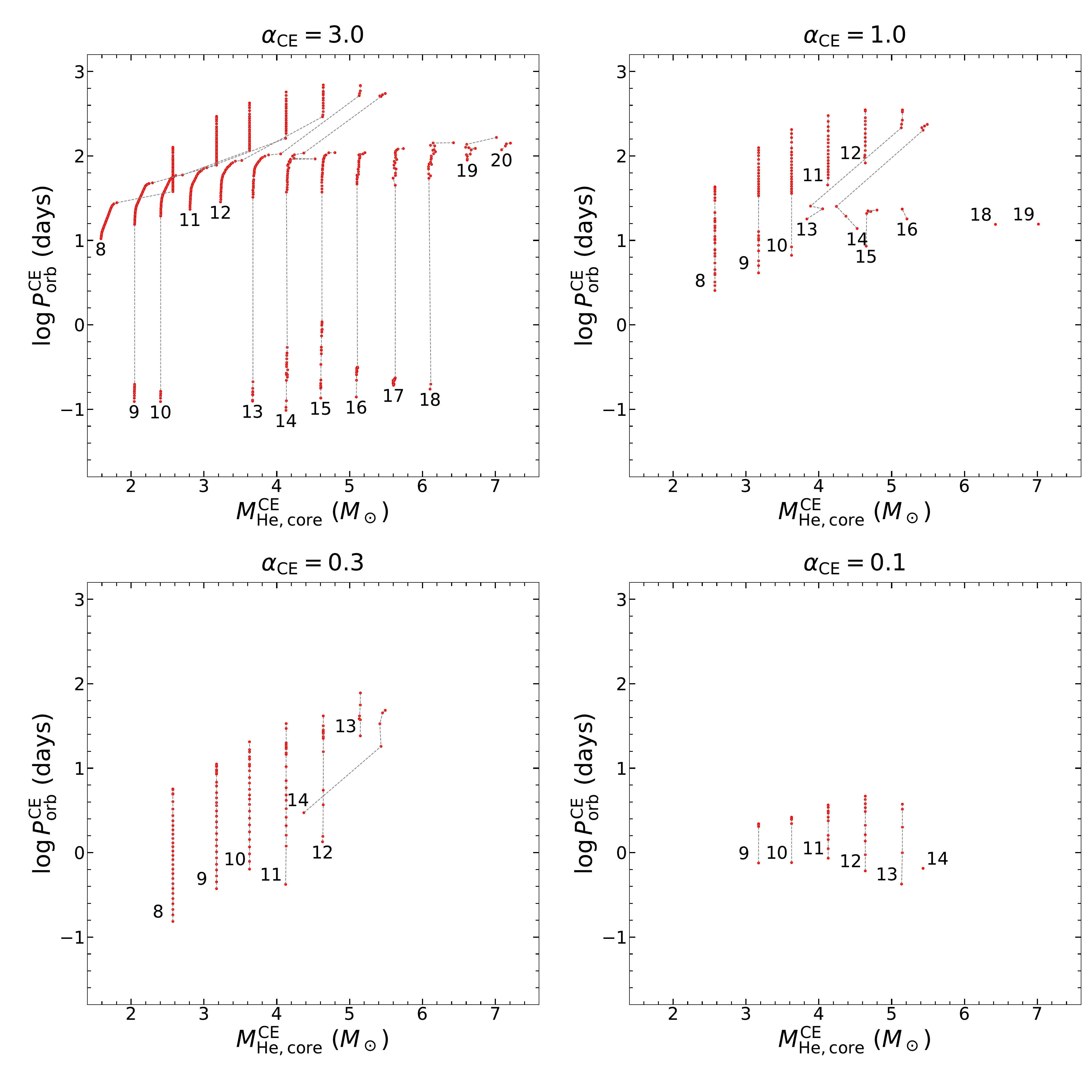}
    \centering
    \caption{Orbital period $\log P^{\rm CE}_{\rm orb}$ as a function of donor's helium core mass $M^{\rm CE}_{\rm He,core}$ for the binaries at the termination of CE evolution, by assuming four different CE ejection efficiencies with $\alpha_{\rm CE}=3.0$, 1.0, 0.3, and 0.1. The number next to each curve gives the initial mass of the donor star adopted in our calculations. The red dots denote all of our calculated systems that survived CE evolution.}
    \label{6}
\end{figure*}

In Figure \ref{4}, we present the distributions of our simulated binaries at the termination of CE evolution in the plane of orbital period $ P_{\rm orb}^{\rm CE}$ versus donor's hydrogen envelope mass $M_{\rm H,env}^{\rm CE}$. The four panels correspond to the calculated results with $\alpha_{\rm CE}=3.0$, 1.0, 0.3, and 0.1, respectively. Overall, the CE survivors with longer $ P_{\rm orb}^{\rm CE}$ are evolved from the initial binaries with longer $ P_{\rm orb}^{\rm i}$, and the orbital period distributions of post-CE binaries are sensitive to the options of $\alpha_{\rm CE}$. There is an obvious tendency that varying $\alpha_{\rm CE}$ from 3.0 to 0.1 can significantly reduce the formation of the systems that survived CE evolution, and $ P_{\rm orb}^{\rm CE}$ can cover a range decreasing from $\sim 0.1-600$ days to  $\sim 0.4-5$ days.  

In the case of $\alpha_{\rm CE}=3.0$, there are no systems with $ P_{\rm orb}^{\rm CE} \sim 1-10$ days. We find that previous CE phases of the systems with $P_{\rm orb}^{\rm CE} \lesssim 1$ day can last about $10^3$~yr, which are much longer than those of the systems with $P_{\rm orb}^{\rm CE} \gtrsim 10$ days (see also Table \ref{tab:t1}). The longer durations of CE phases in the former systems allow the donor stars to eject more hydrogen envelopes and have $M_{\rm H,env}^{\rm CE}\sim 0.3-1.2\,M_\odot $. On the contrary, the donor stars in the latter systems are expected to remain more hydrogen envelopes with $M_{\rm H,env}^{\rm CE}\sim 1.0-4.0\,M_\odot $ after CE evolution. There is an exception for a group of systems with $P_{\rm orb}^{\rm CE} \sim 30-600$ days and $M_{\rm H,env}^{\rm CE}\lesssim 0.4\,M_\odot $, they are post-Case C binaries. As a result, the post-Case B binaries are distributed in two distinct regions with $P_{\rm orb}^{\rm CE} \lesssim 1$ day and $M_{\rm H,env}^{\rm CE}\sim 0.3-1.2\,M_\odot $ or $P_{\rm orb}^{\rm CE} \gtrsim 10$ days and $M_{\rm H,env}^{\rm CE}\sim 1.0-4.0\,M_\odot $. We show in Section \ref{sec:4.3} that the input parameters (i.e., $\dot M_{\rm high}$, $\dot M_{\rm low}$, and $\delta$) of controlling CE evolution are responsible for the formation of these systems with obviously different properties. And, we find that the post-Case B binaries with $P_{\rm orb}^{\rm CE} \lesssim 1$~day do not evolve into CEDPs (see Figure \ref{14} in the Appendix \ref{AB}).

In the cases of $\alpha_{\rm CE} = 1.0$ and $\alpha_{\rm CE} = 0.3$, the systems experienced Case C MT are more likely to survive CE evolution. As a result, most of them having $M_{\rm H,env}^{\rm CE}\lesssim 0.4M_{\rm \odot}$ are post-Case C binaries. And, a small group of systems with $M_{\rm H,env}^{\rm CE}\sim 1.0-2.0M_{\rm \odot}$ are post-Case B binaries. 

In the case of $\alpha_{\rm CE} = 0.1$, all systems survived CE evolution are post-Case C binaries, and the donor stars have relatively low hydrogen envelope masses of $\lesssim 0.2\,M_\odot$.

\subsection{The \texorpdfstring{$M_{\rm H,env}^{\rm f}-\log P_{\rm orb}^{\rm f}$} D Diagram at the Moment of Core Carbon Depletion} \label{sec:3.4}

In Figure \ref{5}, we present the distributions of all binaries that survived CE evolution in the plane of final hydrogen envelope mass $M^{\rm f}_{\rm H,env}$ versus final orbital period $P^{\rm f}_{\rm orb}$. On the one hand, stable MT from more-massive donors to less-massive NSs in CEDPs and Case BB/BC MT phases tends to shrink the binary orbits. On the other hand, the mass loss due to stellar winds from more-massive donors tends to widen the binary orbits when the donors evolve to detach from their Roche lobes. Both of the above processes are able to strip the hydrogen envelopes of the donor stars if remained during CE evolution. On the whole, all CE survivors that experienced Case B MT are expected to contain a donor star without any hydrogen envelope at the moment of core carbon depletion (see also the evolutionary consequence of a post-Case B binary in Section \ref{sec:4.1}). This situation can also apply to some CE survivors that experienced Case C MT (e.g., see Figure \ref{3} involving $\alpha_{\rm CE}=1.0$). In the remaining CE survivors, the donor stars still have $M_{\rm H,env}^{\rm f} \lesssim 0.1-0.2\,M_\odot$ until core carbon depletion (e.g., see Section \ref{sec:4.1} for the evolutionary consequence of a post-Case C binary involving $\alpha_{\rm CE}=3.0$). 

In the case of $\alpha_{\rm CE}=3.0$, the final orbital periods of CE survivors can cover a wide range of $P_{\rm orb}^{\rm f} \sim 0.4-600$~days. All donor stars in post-Case B binaries have $M_{\rm H,env}^{\rm f}=0$ due to the mass transfer in CEDPs and the mass loss via stellar winds \citep{Nugis2000}. 
There are a group of post-Case C binaries with $M_{\rm H,env}^{\rm f}\sim 0.02-0.2\,M_\odot$ and $P_{\rm orb}^{\rm f} \sim 30-600$~days. In the $M_{\rm H,env}^{\rm f}-\log P_{\rm orb}^{\rm f}$ diagram, there exists a distinct turning point for some specific $M_{\rm d}^{\rm i}$. Note that $M_{\rm H,env}^{\rm f}$ depends on the MT rates and the durations of CEDPs, the binaries with longer $P_{\rm orb}^{\rm f}$ are expected to have higher MT rates while shorter MT durations (since the donor stars have evolved to a more advanced stage). As a result, the competition between these two factors leads to the appearance of turning points.

In the cases of $\alpha_{\rm CE}=0.1-1.0$, most CE survivors are post-Case C binaries. The donor stars in some post-Case C binaries also have $M_{\rm H,env}^{\rm f}=0$, which mainly caused by the mass loss via RLOF during CEDPs. For post-Case B binaries, the donor stars do not have any hydrogen envelope at the moment of core carbon depletion. For post-Case C binaries, when decreasing $\alpha_{\rm CE}$ from 1.0 to 0.1, the $P_{\rm orb}^{\rm f}$ distributions cover a range varying from $\sim 1-400$~days to $\sim 0.4-4$~days, and the maximum $M_{\rm H,env}^{\rm f}$ decreases from $\sim 0.2\,M_\odot$ to $\sim 0.1\,M_\odot$. 


\begin{figure*}[htbp]
    \includegraphics[width=0.9\textwidth]{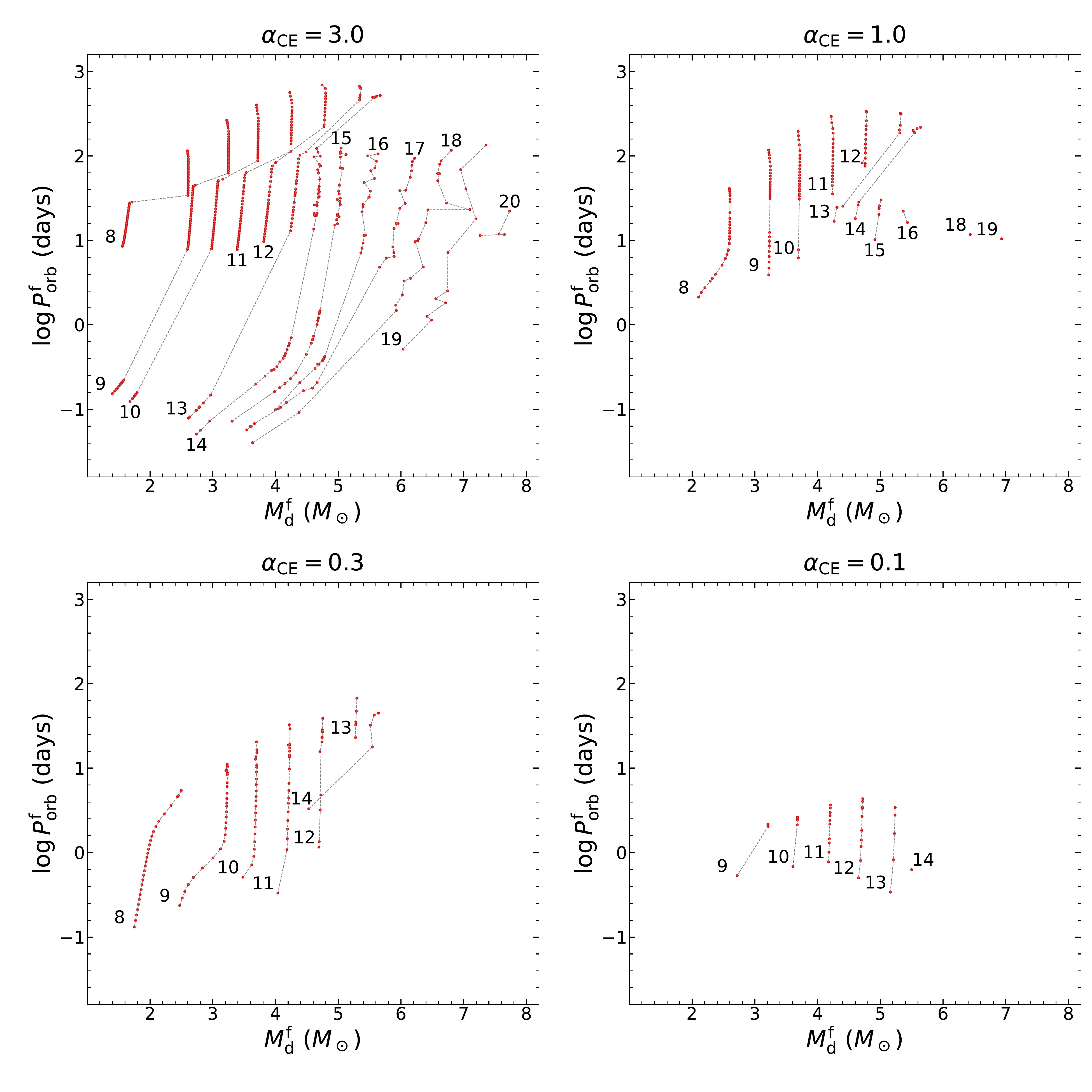}
    \centering
    \caption{Final orbital period $\log P^{\rm f}_{\rm orb}$ as a function of final donor mass $M^{\rm f}_{\rm d}$ for CE survivors, by assuming four different CE ejection efficiencies with $\alpha_{\rm CE}=3.0$, 1.0, 0.3, and 0.1. The number next to each curve gives the initial mass of the donor star adopted in our calculations. The red dots denote all of our calculated systems that survived CE evolution.} 
    \label{7}
\end{figure*}

\subsection{The \texorpdfstring{$M^{\rm CE}_{\rm He,core}-\log P^{\rm CE}_{\rm orb}$} D Diagram at the Termination of CE Evolution}

Figure \ref{6} shows the distributions of our simulated binaries at the termination of CE evolution in the plane of orbital period $P_{\rm orb}^{\rm CE}$ versus donor's helium core mass $M_{\rm He,core}^{\rm CE}$, by assuming four different CE ejection efficiencies with $\alpha_{\rm CE}=3.0$, 1.0, 0.3, and 0.1. Here, $M_{\rm He,core}^{\rm CE}$ includes the contribution from  the carbon-oxygen core and its helium envelope if a carbon-oxygen core has been developed.  We can see that the binaries survived CE evolution always have the donor stars with $M_{\rm He,core}^{\rm CE}\lesssim 7\,M_\odot$.  

Evolved from the initial binaries with a specific $M_{\rm d}^{\rm i}$, the CE survivors experienced Case C MT have relatively long orbital periods and the donor stars have almost the same $M_{\rm He,core}^{\rm CE}$. This $M_{\rm He,core}^{\rm CE}$ is $\sim 1\,M_{\odot}$ more massive than those for the donor stars in post-Case B binaries. Overall, the $\alpha_{\rm CE}=0.1$ case only allows the formation of the CE survivors with a narrow range of $M_{\rm He,core}^{\rm CE}\sim 3.2-5.4\,M_\odot$, compared to $M_{\rm He,core}^{\rm CE}\sim 1.6-7.2\,M_\odot$ in the $\alpha_{\rm CE}=3.0$ case.

\begin{figure*}[htbp]
    \includegraphics[width=0.9\textwidth]{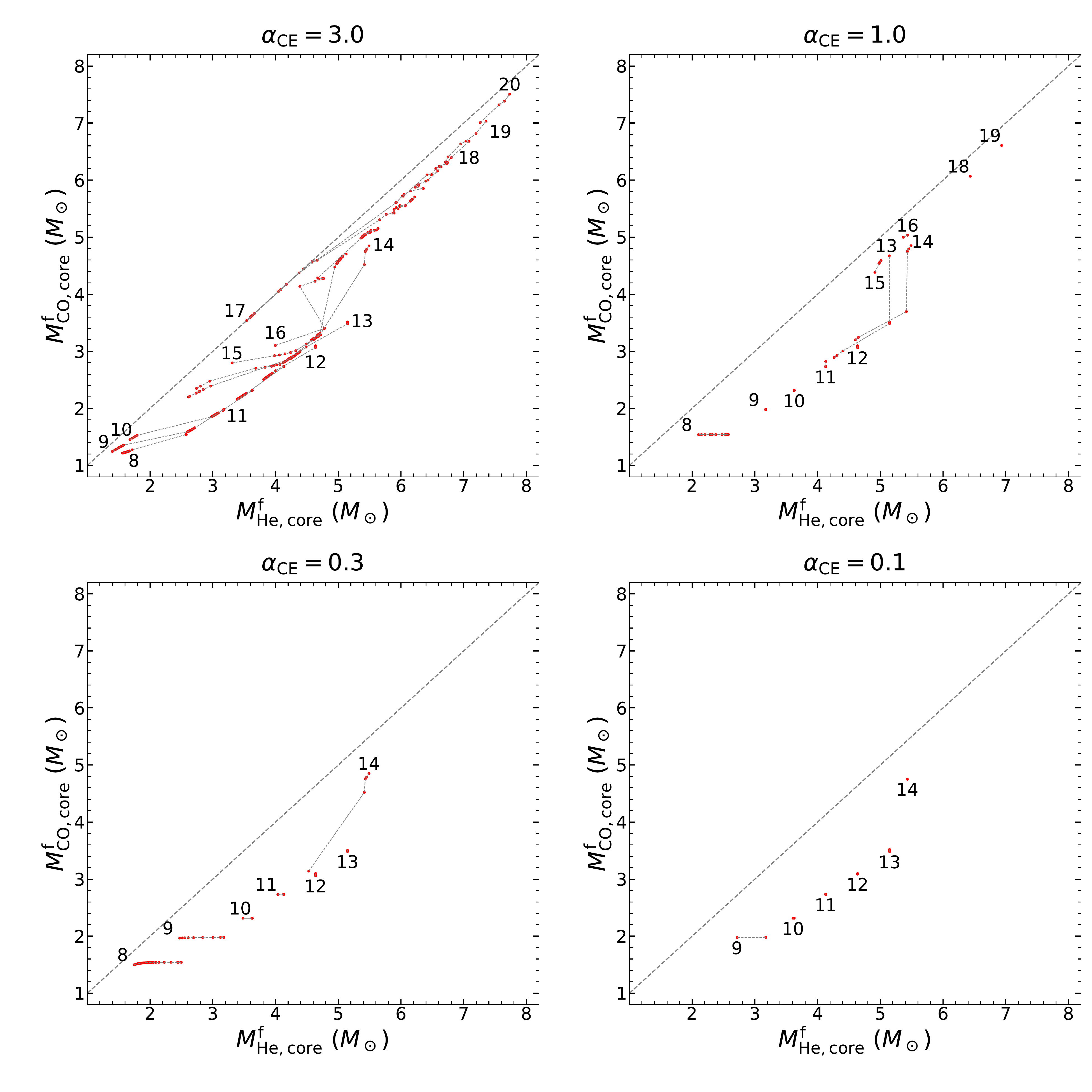}
    \centering
    \caption{Final carbon-oxygen core mass $M^{\rm f}_{\rm CO,core}$ as a function of final helium core mass $M^{\rm f}_{\rm He,core}$ for the donor stars in CE survivors, by assuming four different CE ejection efficiencies with $\alpha_{\rm CE}=3.0$, 1.0, 0.3, and 0.1. In each panel, the dashed line represents the relation of $M^{\rm f}_{\rm He,core}=M^{\rm f}_{\rm CO,core}$. The number next to each curve gives the initial mass of the donor star adopted in our calculations. The red dots denote all of our calculated systems that survived CE evolution. }
    \label{8}
\end{figure*}

Cygnus X-3 is an X-ray binary containing a Wolf-Rayet star and a compact object in a $\sim 0.2$-day orbit \citep{vanKerkwijk1992}. The masses of the Wolf-Rayet star and the compact object are estimated to be $10.3^{+3.9}_{-2.8}\,M_\odot$ and $2.4^{+2.1}_{-1.1}\,M_\odot$, respectively \citep{Zdziarski2013}. The mass range of the compact object allows it to be either an NS or a black hole. Recent observations suggest the compact object to be a black hole, but the possibility of an NS cannot be ruled out \citep{Antokhin2022,Ge2024}. Based on our simulations, the helium core masses at the termination of CE evolution are always less than $\sim 7M_{\rm \odot}$ even in the case of $\alpha_{\rm CE} \geq 1.0$. Hence, we propose that the compact object in Cygnus X-3 is more likely to be a low-mass black hole \citep[see also][]{Wang2024}. 

\subsection{The \texorpdfstring{$M^{\rm f}_{\rm d}-\log P^{\rm f}_{\rm orb}$} D Diagram at the Moment of Core Carbon Depletion}\label{Sec:3.6}

\begin{figure*}[htbp]
    \includegraphics[width=0.9\textwidth]{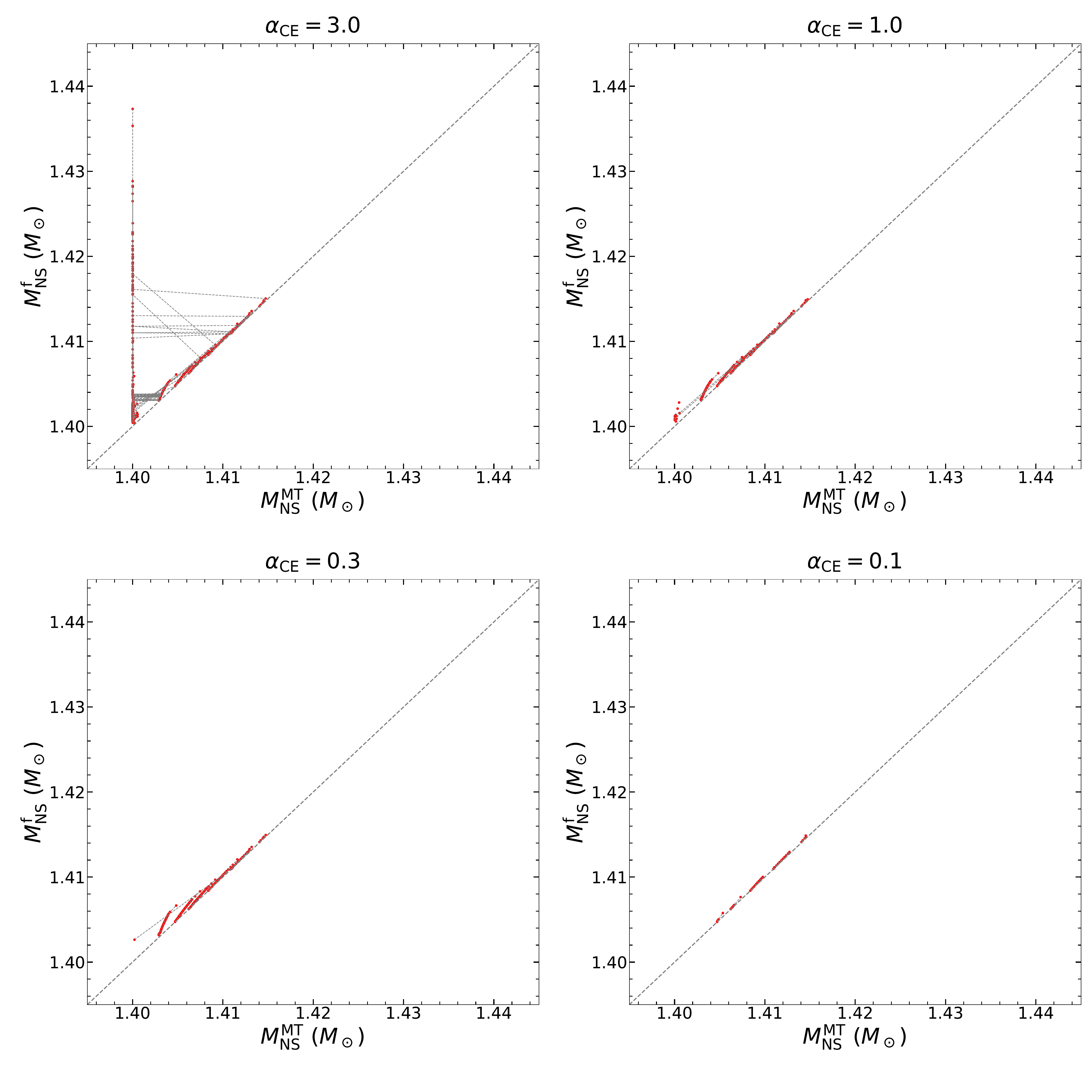}
    \centering
    \caption{Final NS mass $M^{\rm f}_{\rm NS}$ as a function of the NS mass $M^{\rm MT}_{\rm NS}$ at the onset of the first RLOF MT, by assuming four different CE ejection efficiencies with $\alpha_{\rm CE}=3.0$, 1.0, 0.3, and 0.1. In each panel, the dashed line represents the relation of $M^{\rm f}_{\rm NS}=M^{\rm MT}_{\rm NS}$. The red dots show all of our calculated systems that survived CE evolution.}
    \label{9}
\end{figure*}

Figure \ref{7} presents the distributions of our simulated binaries survived CE evolution in the plane of final orbital period $P_{\rm orb}^{\rm f}$ versus final donor mass $M_{\rm d}^{\rm f}$. For a specific $M_{\rm d}^{\rm i}$, the donor stars in post-Case C binaries with relatively long $P_{\rm orb}^{\rm f}$  have almost the same $M_{\rm d}^{\rm f}$, which is $\gtrsim 1M_{\rm \odot}$ more massive than those for the donor stars in post-Case B binaries with relatively short $P_{\rm orb}^{\rm f}$.
Our simulation indicates that the minimum $P_{\rm orb}^{\rm f}$ are about 0.04 days in the $\alpha_{\rm CE}=3.0$ case, 2.1 days in the $\alpha_{\rm CE}=1.0$ case, 0.1 days in the $\alpha_{\rm CE}=0.3$ case, and 0.4 days in the $\alpha_{\rm CE}=0.1$ case. Therefore, these minimum $P_{\rm orb}^{\rm f}$ are sensitive to the options of $\alpha_{\rm CE}$ (see also the discussion on the parameter spaces of CE survivors in Section \ref{sec:3.1}). 

Considering the dynamical effect of supernovae in binaries, \citet{Tauris2017} used the observed properties of Galactic DNSs to constrain the orbital periods of the binaries with an NS and a pre-supernova helium star. For the formation of some close DNSs with orbital periods of $\sim 0.1$ days \citep[e.g., PSR J0737-3039 in Figure 25 of][]{Tauris2017}, it is required that the pre-supernova binaries are also close systems with $P_{\rm orb}^{\rm f}\sim 0.1$ days.  Thus, it seems that the observations of close DNS systems favor the assumptions of $\alpha_{\rm CE}=3.0$ and $\alpha_{\rm CE}=0.3$. It has been pointed out that the donor stars in HMXBs have a rejuvenated structure due to mass accretion during HMXB's progenitor evolution \citep{Landri2024}, and the NSs are likely to accrete a modest amount of matter \citep{MacLeod2015a,MacLeod2015} along with the launch of jets \citep{Papish2015,Shiber2019} during a CE inspiral phase. Our simulations do not include these processes, which can potentially change the configuration of CE evolution.

\subsection{The \texorpdfstring{$M^{\rm f}_{\rm He,core}-M^{\rm f}_{\rm CO,core}$} D Diagram at the Moment of Core Carbon Depletion}

In Figure \ref{8}, we present the distribution of all the binaries survived CE evolution in the plane of final helium core mass $M_{\rm He,core}^{\rm f}$ versus final carbon-oxygen core mass $M_{\rm CO,core}^{\rm f}$. We obtain from our simulation that $M_{\rm CO,core}^{\rm f}$ of the donor stars in post-Case B binaries is $\sim 1-2\,M_{\odot}$ less massive than those of the donor stars in post-Case C binaries. For a specific $M_{\rm d}^{\rm i}$, the donor stars in post-Case C binaries have almost the same $M_{\rm CO,core}^{\rm f}$. Generally, $M_{\rm CO,core}^{\rm f}$ can cover a broad range from $1.2\,M_{\odot}$
to $7.5\,M_{\odot}$ across all of our adopted $\alpha_{\rm CE}$. 

Ultra-stripped supernovae are related to the pre-supernova objects with helium envelope masses of $\lesssim 0.2\,M_\odot$ due to extreme stripping in binaries \citep{Tauris2015}. Using $M_{\rm He,env}^{\rm f}$ ($=M^{\rm f}_{\rm He,core}-M^{\rm f}_{\rm CO,core} $) to represent the masses of helium envelope at the moment of core carbon depletion, we check whether our simulation is able to produce such objects with $M_{\rm He,env}^{\rm f} \lesssim 0.2\,M_\odot$.

In the case of $\alpha_{\rm CE}=3.0$, for the post-Case B binaries with $M_{\rm d}^{\rm i}\geq 14\,M_{\odot}$, stellar winds \citep{Nugis2000} play a leading role in the stripping of donor's helium envelopes. 
In these systems, the donor stars have $M_{\rm He,env}^{\rm f}\lesssim 2\,M_{\odot}$. For some binaries with $M_{\rm d}^{\rm i}=17\,M_{\odot}$ or $M_{\rm d}^{\rm i}=18\,M_{\odot}$, in addition to stellar winds, the process of Case BB/BC MT further strips away the remaining helium envelopes of the donor stars. As a consequence, the final donors have $M_{\rm He,env}^{\rm f}=0$. For the post-Case B binaries with $M_{\rm d}^{\rm i}= 8-13\,M_{\odot}$, the process of Case BB/BC MT dominates the stripping of the helium envelopes of the donor stars in some close systems, probably leading to the formation of ultra-stripped supernovae from the exploding donor stars with $M_{\rm He,env}^{\rm f}\sim 0.2M_{\rm \odot}$. In wide systems, the final donor stars have $M_{\rm He,env}^{\rm f}\sim 1.0M_{\rm \odot}$. 
For post-Case C binaries, the donor stars are expected to have $M_{\rm He,env}^{\rm f}\sim 0.5-1.7M_{\rm \odot}$. This result means that ultra-stripped supernovae will not happen in post-Case C binaries.

In the cases of $\alpha_{\rm CE}=0.1-1.0$, less binaries are able to survive CE evolution compared to the case of $\alpha_{\rm CE}=3.0$. The donor stars have $M_{\rm He,env}^{\rm f}\sim 0.2-1.0\,M_{\odot}$ in post-Case B binaries and $M_{\rm He,env}^{\rm f}\sim 0.3-1.7\,M_{\odot}$ in post-Case C binaries.

\subsection{The \texorpdfstring{$M^{\rm f}_{\rm NS}-M^{\rm MT}_{\rm NS}$} D Diagram}

In Figure \ref{9}, we present the distribution of all binaries survived CE evolution in the plane of final NS mass $M^{\rm f}_{\rm NS}$ versus the NS mass $M^{\rm MT}_{\rm NS}$ at the onset of RLOF MT. For clarity, we did not plot $M_{\rm d}^{\rm i}$ next to each curve in this diagram. Note that the initial NS mass is set to be $1.4\,M_\odot$. On the whole, the maximum accreted mass of NSs is $\sim 0.04M_{\rm \odot}$ across all our adopted $\alpha_{\rm CE}$. The NSs in post-Case C binaries can accrete $\sim 0.01M_{\rm \odot}$ matter through a stellar wind \citep{deJager1988} before the onset of RLOF MT. During the CEDPs which last $\lesssim 0.1$~Myr, the NSs can accrete $\sim 0.001M_{\rm \odot}$. 

The NSs in post-Case B binaries can hardly accrete matter via the \citet{deJager1988} wind before the onset of RLOF MT. In the case of $\alpha_{\rm CE}=3.0$, for the post-Case B binaries with $M_{\rm d}^{\rm i}\geq 9\,M_{\odot}$, the NSs mainly accrete matter through a stellar wind \citep{Nugis2000} after CE evolution. For the systems with $M_{\rm d}^{\rm i}=8\,M_{\odot}$, the mass increase of NSs via RLOF in CEDPs and Case BB/BC MT phases is comparable to that via capturing donor's wind (see also Figure \ref{11}). Overall, the NSs accrete $\sim 0.01-0.04M_{\rm \odot}$ in close post-Case B binaries and $\lesssim 0.01M_{\rm \odot}$ in wide systems. 
In the cases of $\alpha_{\rm CE}=0.1-1.0$, a small fraction of CE survivors are post-Case B binaries in which the NSs accrete material primarily through the process of capturing  a stellar wind \citep{Nugis2000}. 

\section{Discussion} \label{sec:discussion}
\begin{figure*}[htbp]
    \includegraphics[width=0.9\textwidth]{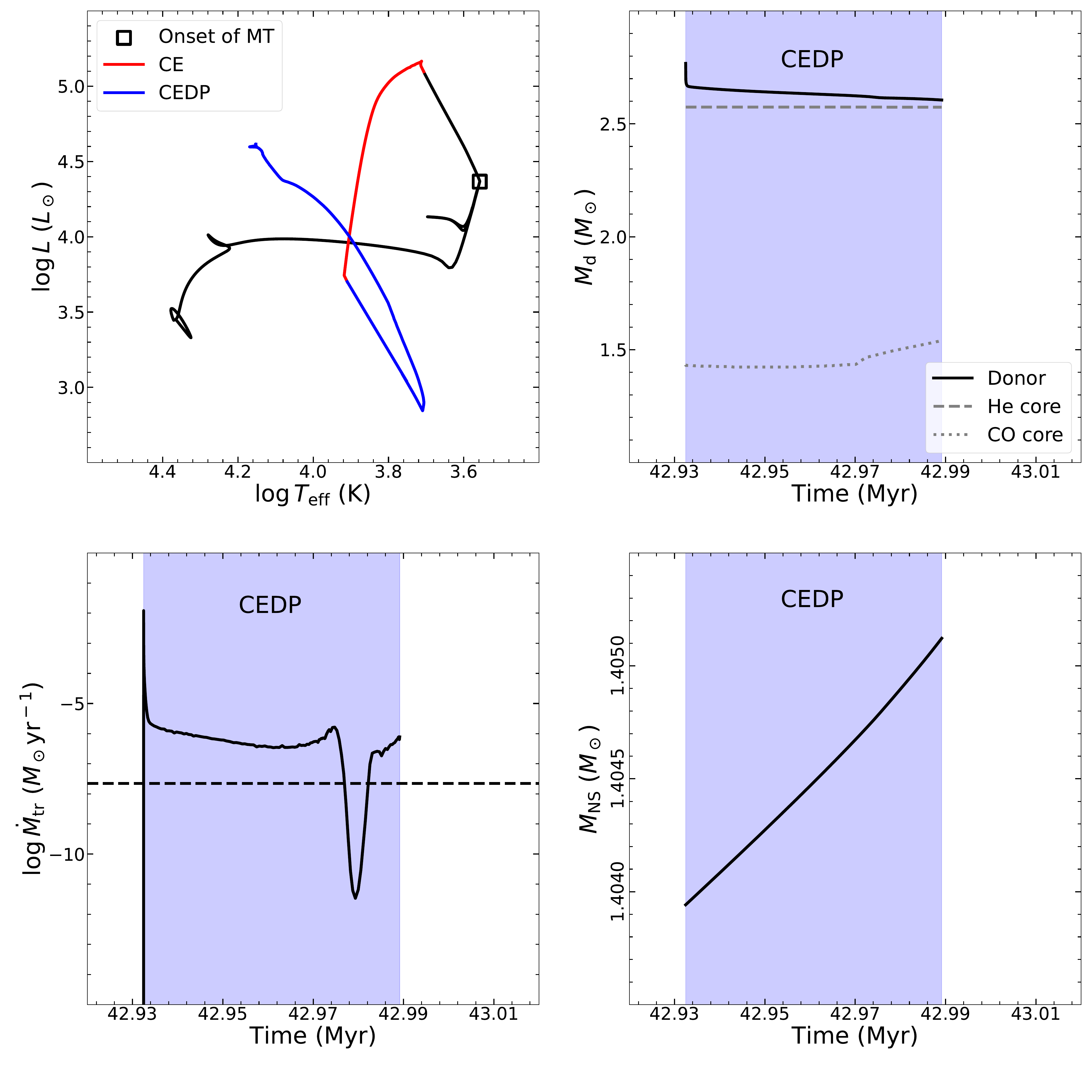}
    \centering
    \caption{Similar to Figure \ref{3}, but adopting $\alpha_{\rm CE}=3.0$ for CE evolution.}
    \label{10}
\end{figure*}

\begin{figure*}[htbp]
    \includegraphics[width=0.9\textwidth]{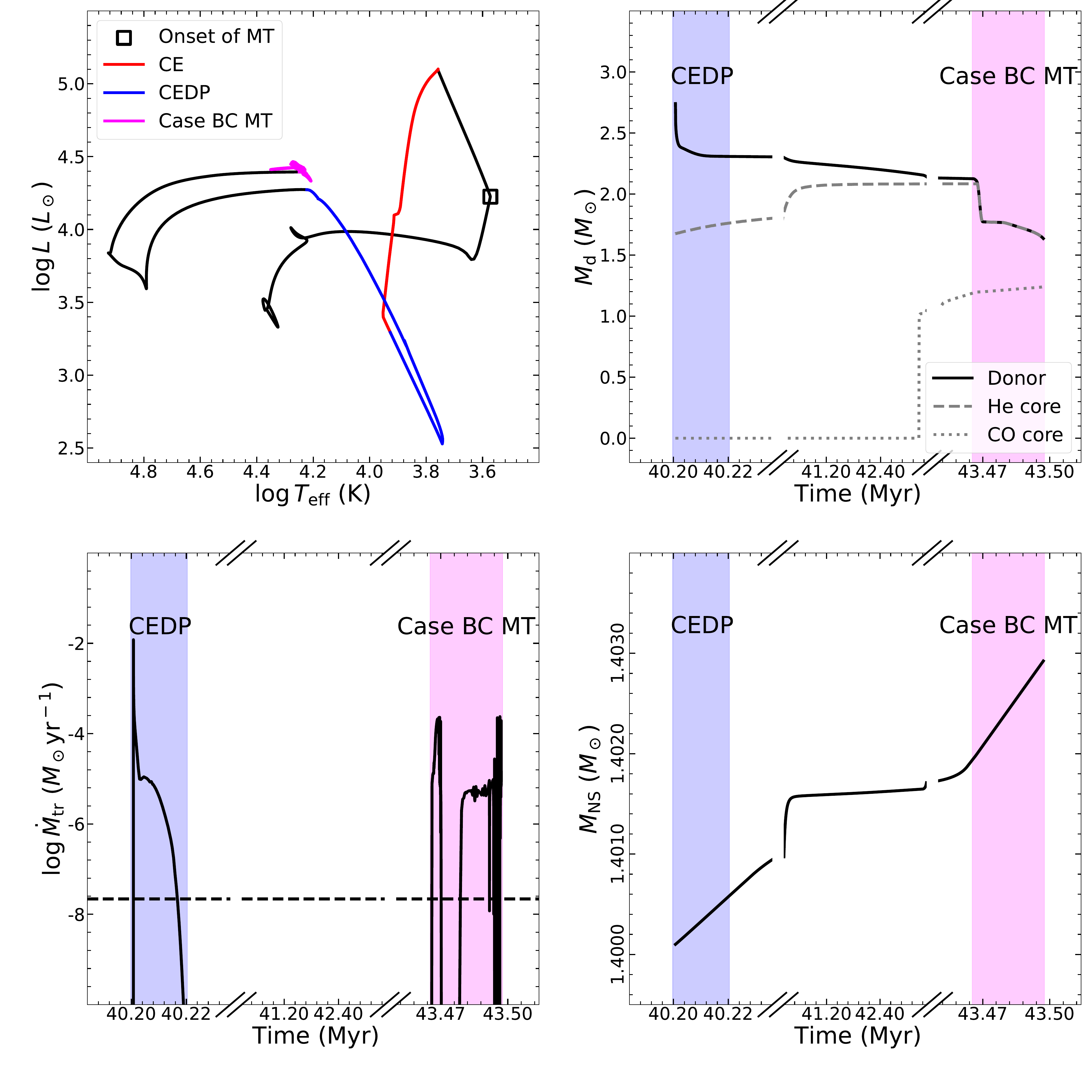}
    \centering
    \caption{Evolutionary tracks for the binary with a $1.4M_{\odot}$ NS and a $8M_{\odot}$ donor in a 537-day orbit. Here, $\alpha_{\rm CE}=3.0$ is adopted. Upper left panel: Hertzsprung-Russell diagram for the donor star. We mark the position at the onset of mass transfer with the symbol of a black square. The red and blue solid curves correspond to the binary undergoing CE and CEDP, respectively. The magenta solid curve represents the Case BC MT stage. Upper right panel: Mass of the donor star/the helium core/the carbon-oxygen core as a function of time after the binary finished CE evolution. Lower left panel: Evolution of RLOF MT rate as a function of time. The black dashed line  represents the Eddington  limit. Lower right panel: Mass of the NS as a function of time.}
    \label{11}
\end{figure*}

\begin{table*}[htbp]
\centering
\begin{tabular}{m{3em}m{5em}m{5.5em}m{6em}m{6em}m{4em}m{5.5em}m{6em}m{5em}}
\hline
\hline
\multirow{1}{=}{\centering $\alpha_{\rm CE}$} &\multirow{1}{=}{\centering Model} &\multirow{1}{=}{\centering $\log (P_{\rm orb}^{\rm i}/\rm d)$} &\multirow{1}{=}{\centering Duration$\,$[yr]} &\multirow{1}{=}{\centering $M_{\rm H,env}^{ \mathcal{CE} }$$\,$[$M_{\rm \odot}$]} &\multirow{1}{=}{\centering $P_{\rm orb}^{\mathcal{CE} }$$\,$[days]} &\multirow{1}{=}{\centering $-E_{\rm bind}$$\,$[erg]} &\multirow{1}{=}{\centering $M_{\rm H,env}^{\rm CE}$$\,$[$M_{\rm \odot}$]} &\multirow{1}{=}{\centering $P_{\rm orb}^{\rm CE}$$\,$[days]}\\
\hline
\multirow{14}{=}{\centering 3.0} &\multirow{2}{=}{\centering Default} &\multirow{1}{=}{\centering 2.75} &\multirow{1}{=}{\centering 1761} &\multirow{1}{=}{\centering 7.10} &\multirow{1}{=}{\centering 409.31} &\multirow{1}{=}{\centering $1.95\times 10^{\rm 49}$} &\multirow{1}{=}{\centering 0.45} &\multirow{1}{=}{\centering 0.16}\\
&&\multirow{1}{=}{\centering 2.76} &\multirow{1}{=}{\centering 5} &\multirow{1}{=}{\centering 7.10} &\multirow{1}{=}{\centering 418.38} &\multirow{1}{=}{\centering $1.95\times 10^{\rm 49}$} &\multirow{1}{=}{\centering 2.05} &\multirow{1}{=}{\centering 19.46}\\
\cline{2-9}
&\multirow{2}{=}{\centering $4\dot M_{\rm high}$} &\multirow{1}{=}{\centering 2.75} &\multirow{1}{=}{\centering 1687} &\multirow{1}{=}{\centering 7.05} &\multirow{1}{=}{\centering 363.01} &\multirow{1}{=}{\centering $1.95\times 10^{\rm 49}$} &\multirow{1}{=}{\centering 0.45} &\multirow{1}{=}{\centering 0.16}\\
&&\multirow{1}{=}{\centering 2.76} &\multirow{1}{=}{\centering 1} &\multirow{1}{=}{\centering 7.05} &\multirow{1}{=}{\centering 371.58} &\multirow{1}{=}{\centering $1.95\times 10^{\rm 49}$} &\multirow{1}{=}{\centering 2.05} &\multirow{1}{=}{\centering 19.04}\\
&\multirow{2}{=}{\centering $0.25\dot M_{\rm high}$} &\multirow{1}{=}{\centering 2.75} &\multirow{1}{=}{\centering 1770} &\multirow{1}{=}{\centering 7.14} &\multirow{1}{=}{\centering 441.73} &\multirow{1}{=}{\centering $1.95\times 10^{\rm 49}$} &\multirow{1}{=}{\centering 0.45} &\multirow{1}{=}{\centering 0.16}\\
&&\multirow{1}{=}{\centering 2.76} &\multirow{1}{=}{\centering 20} &\multirow{1}{=}{\centering 7.14} &\multirow{1}{=}{\centering 450.19} &\multirow{1}{=}{\centering $1.95\times 10^{\rm 49}$} &\multirow{1}{=}{\centering 2.03} &\multirow{1}{=}{\centering 18.88}\\
\cline{2-9}
&\multirow{2}{=}{\centering $4\dot M_{\rm low}$} &\multirow{1}{=}{\centering 2.75} &\multirow{1}{=}{\centering 1560} &\multirow{1}{=}{\centering 7.10} &\multirow{1}{=}{\centering 409.31} &\multirow{1}{=}{\centering $1.95\times 10^{\rm 49}$} &\multirow{1}{=}{\centering 0.45} &\multirow{1}{=}{\centering 0.16}\\
&&\multirow{1}{=}{\centering 2.76} &\multirow{1}{=}{\centering 1623} &\multirow{1}{=}{\centering 7.10} &\multirow{1}{=}{\centering 418.38} &\multirow{1}{=}{\centering $1.95\times 10^{\rm 49}$} &\multirow{1}{=}{\centering 0.47} &\multirow{1}{=}{\centering 0.17}\\
&\multirow{2}{=}{\centering $0.25\dot M_{\rm low}$} &\multirow{1}{=}{\centering 2.75} &\multirow{1}{=}{\centering 5} &\multirow{1}{=}{\centering 7.10} &\multirow{1}{=}{\centering 409.31} &\multirow{1}{=}{\centering $1.95\times 10^{\rm 49}$} &\multirow{1}{=}{\centering 2.05} &\multirow{1}{=}{\centering 18.41}\\
&&\multirow{1}{=}{\centering 2.76} &\multirow{1}{=}{\centering 5} &\multirow{1}{=}{\centering 7.10} &\multirow{1}{=}{\centering 418.38} &\multirow{1}{=}{\centering $1.95\times 10^{\rm 49}$} &\multirow{1}{=}{\centering 2.05} &\multirow{1}{=}{\centering 19.49}\\
\cline{2-9}
&\multirow{2}{=}{\centering $\delta=0.001$} &\multirow{1}{=}{\centering 2.75} &\multirow{1}{=}{\centering 5} &\multirow{1}{=}{\centering 7.10} &\multirow{1}{=}{\centering 409.31} &\multirow{1}{=}{\centering $1.95\times 10^{\rm 49}$} &\multirow{1}{=}{\centering 2.06} &\multirow{1}{=}{\centering 18.55}\\
&&\multirow{1}{=}{\centering 2.76} &\multirow{1}{=}{\centering 5} &\multirow{1}{=}{\centering 7.10} &\multirow{1}{=}{\centering 418.38} &\multirow{1}{=}{\centering $1.95\times 10^{\rm 49}$} &\multirow{1}{=}{\centering 2.05} &\multirow{1}{=}{\centering 19.60}\\
&\multirow{2}{=}{\centering $\delta=0.04$}&\multirow{1}{=}{\centering 2.75} &\multirow{1}{=}{\centering 1899} &\multirow{1}{=}{\centering 7.10} &\multirow{1}{=}{\centering 409.31} &\multirow{1}{=}{\centering $1.95\times 10^{\rm 49}$} &\multirow{1}{=}{\centering 0.41} &\multirow{1}{=}{\centering 0.15}\\
&&\multirow{1}{=}{\centering 2.76} &\multirow{1}{=}{\centering 1803} &\multirow{1}{=}{\centering 7.10} &\multirow{1}{=}{\centering 418.38} &\multirow{1}{=}{\centering $1.95\times 10^{\rm 49}$} &\multirow{1}{=}{\centering 0.43} &\multirow{1}{=}{\centering 0.16}\\
\hline
\multirow{14}{=}{\centering 1.0} &\multirow{2}{=}{\centering Default} &\multirow{1}{=}{\centering 3.07} &\multirow{1}{=}{\centering 214} &\multirow{1}{=}{\centering 4.83} &\multirow{1}{=}{\centering 961.89} &\multirow{1}{=}{\centering $3.68\times 10^{\rm 48}$} &\multirow{1}{=}{\centering 0.15} &\multirow{1}{=}{\centering 8.37}\\
&&\multirow{1}{=}{\centering 3.08} &\multirow{1}{=}{\centering 5} &\multirow{1}{=}{\centering 4.83} &\multirow{1}{=}{\centering 984.64} &\multirow{1}{=}{\centering $3.48\times 10^{\rm 48}$} &\multirow{1}{=}{\centering 0.22} &\multirow{1}{=}{\centering 36.00}\\
\cline{2-9}
&\multirow{2}{=}{\centering $4\dot M_{\rm high}$} &\multirow{1}{=}{\centering 3.07} &\multirow{1}{=}{\centering 1} &\multirow{1}{=}{\centering 4.77} &\multirow{1}{=}{\centering 880.09} &\multirow{1}{=}{\centering $3.68\times 10^{\rm 48}$} &\multirow{1}{=}{\centering 0.24} &\multirow{1}{=}{\centering 34.55}\\
&&\multirow{1}{=}{\centering 3.08} &\multirow{1}{=}{\centering 1} &\multirow{1}{=}{\centering 4.77} &\multirow{1}{=}{\centering 901.91} &\multirow{1}{=}{\centering $3.48\times 10^{\rm 48}$} &\multirow{1}{=}{\centering 0.23} &\multirow{1}{=}{\centering 37.09}\\
&\multirow{2}{=}{\centering $0.25\dot M_{\rm high}$} &\multirow{1}{=}{\centering 3.07} &\multirow{1}{=}{\centering 227} &\multirow{1}{=}{\centering 4.87} &\multirow{1}{=}{\centering 1001.01} &\multirow{1}{=}{\centering $3.68\times 10^{\rm 48}$} &\multirow{1}{=}{\centering 0.15} &\multirow{1}{=}{\centering 8.30}\\
&&\multirow{1}{=}{\centering 3.08} &\multirow{1}{=}{\centering 19811} &\multirow{1}{=}{\centering 4.87} &\multirow{1}{=}{\centering 1039.79} &\multirow{1}{=}{\centering $3.48\times 10^{\rm 48}$} &\multirow{1}{=}{\centering 0.07} &\multirow{1}{=}{\centering 9.33}\\
\cline{2-9}
&\multirow{2}{=}{\centering $4\dot M_{\rm low}$} &\multirow{1}{=}{\centering 3.07} &\multirow{1}{=}{\centering 178} &\multirow{1}{=}{\centering 4.83} &\multirow{1}{=}{\centering 961.89} &\multirow{1}{=}{\centering $3.68\times 10^{\rm 48}$} &\multirow{1}{=}{\centering 0.15} &\multirow{1}{=}{\centering 8.76}\\
&&\multirow{1}{=}{\centering 3.08} &\multirow{1}{=}{\centering 33} &\multirow{1}{=}{\centering 4.83} &\multirow{1}{=}{\centering 984.64} &\multirow{1}{=}{\centering $3.48\times 10^{\rm 48}$} &\multirow{1}{=}{\centering 0.16} &\multirow{1}{=}{\centering 17.79}\\
&\multirow{2}{=}{\centering $0.25\dot M_{\rm low}$} &\multirow{1}{=}{\centering 3.07} &\multirow{1}{=}{\centering 5} &\multirow{1}{=}{\centering 4.83} &\multirow{1}{=}{\centering 961.89} &\multirow{1}{=}{\centering $3.68\times 10^{\rm 48}$} &\multirow{1}{=}{\centering 0.23} &\multirow{1}{=}{\centering 33.80}\\
&&\multirow{1}{=}{\centering 3.08} &\multirow{1}{=}{\centering 5} &\multirow{1}{=}{\centering 4.83} &\multirow{1}{=}{\centering 984.64} &\multirow{1}{=}{\centering $3.48\times 10^{\rm 48}$} &\multirow{1}{=}{\centering 0.22} &\multirow{1}{=}{\centering 36.01}\\
\cline{2-9}
&\multirow{2}{=}{\centering $\delta=0.001$} &\multirow{1}{=}{\centering 3.07} &\multirow{1}{=}{\centering 5} &\multirow{1}{=}{\centering 4.83} &\multirow{1}{=}{\centering 961.89} &\multirow{1}{=}{\centering $3.68\times 10^{\rm 48}$} &\multirow{1}{=}{\centering 0.23} &\multirow{1}{=}{\centering 33.93}\\
&&\multirow{1}{=}{\centering 3.08} &\multirow{1}{=}{\centering 5} &\multirow{1}{=}{\centering 4.83} &\multirow{1}{=}{\centering 984.64} &\multirow{1}{=}{\centering $3.48\times 10^{\rm 48}$} &\multirow{1}{=}{\centering 0.23} &\multirow{1}{=}{\centering 36.20}\\
&\multirow{2}{=}{\centering $\delta=0.04$}&\multirow{1}{=}{\centering 3.07} &\multirow{1}{=}{\centering 225} &\multirow{1}{=}{\centering 4.83} &\multirow{1}{=}{\centering 961.89} &\multirow{1}{=}{\centering $3.68\times 10^{\rm 48}$} &\multirow{1}{=}{\centering 0.15} &\multirow{1}{=}{\centering 8.21}\\
&&\multirow{1}{=}{\centering 3.08} &\multirow{1}{=}{\centering 160} &\multirow{1}{=}{\centering 4.83} &\multirow{1}{=}{\centering 984.64} &\multirow{1}{=}{\centering $3.48\times 10^{\rm 48}$} &\multirow{1}{=}{\centering 0.15} &\multirow{1}{=}{\centering 9.97}\\
\hline
\hline
\vspace{2mm}
\end{tabular}
\caption{Different evolutionary consequences for some specific binaries with $M_{\rm d}^{\rm i}=10\,M_{\odot}$, when changing the input parameters of controlling CE evolution in the MESA code. The default input parameters are set as $\dot M_{\rm high}=1\,M_{\rm \odot}\, \rm yr^{\rm -1}$, $\dot M_{\rm low}=10^{\rm -5}\,M_{\rm \odot}\, \rm yr^{\rm -1}$ and $\delta =0.02$. Column 4 means the duration that a CE phase lasts. Columns 5 and 6 mean the hydrogen envelope mass $M_{\rm H,env}^{\mathcal{CE} }$ and the orbital period $P_{\rm orb}^{\mathcal{CE} }$ for the binary at the moment when CE evolution is triggered, respectively.}
\label{tab:t1}
\end{table*}

\subsection{The Effect of CE ejection efficiencies on CEDPs} \label{sec:4.1}

In this section, we discuss the effect of $\alpha_{\rm CE}$ on CEDPs. Figure \ref{10} shows the evolutionary tracks of the same binary as in Figure \ref{3}, but $\alpha_{\rm CE}=3.0$ is adopted.
At the time of $\sim 42.93$~Myr, $\sim 0.2\,M_{\odot}$ hydrogen envelope is remained after CE evolution. At the moment of core carbon depletion, most of the hydrogen envelope has been stripped away in the CEDP and the donor star has $M_{\rm H,env}^{\rm f}\sim 0.03\,M_{\odot}$. During the CEDP, the NS accretes $\sim 0.001M_{\rm \odot}$ matter. The final binary has the orbital period of $P_{\rm orb}^{\rm f}\sim 39$ days, which is greatly larger than that ($\sim 3.3$ days) in Figure \ref{3} with $\alpha_{\rm CE}=1.0$. Adopting different $\alpha_{\rm CE}$ results in a big difference between the remaining envelope masses for the final donors (i.e., $M_{\rm H,env}^{\rm f}\sim 0.03\,M_{\odot}$ and $M_{\rm He,env}^{\rm f}\sim 1.1\,M_{\odot}$ in the $\alpha_{\rm CE}=3.0$ case while $M_{\rm H,env}^{\rm f}\sim 0$ and $M_{\rm He,env}^{\rm f}\sim 0.8\,M_{\odot}$ in the $\alpha_{\rm CE}=1.0$ case).

Figure \ref{11} shows the evolutionary tracks of the initial binary with the same component masses in a narrower orbit of $P_{\rm orb}^{\rm i} = 537$ days. Also, we adopt $\alpha_{\rm CE}=3.0$. This binary represents a post-CE survivor experienced Case B MT. At the time of $\sim 40.20\, \rm Myr$, CE evolution finishes and the donor star has a radius of $\sim 25\,R_{\odot}$. At the time of $\sim 40.22\, \rm Myr$, the donor star shrinks quickly until its radius is less than $1\,R_{\odot}$ and detaches from its Roche-lobe. During this CEDP, the donor star loses $\sim 0.5\,M_{\odot}$ hydrogen envelope and the NS accretes $\sim 0.001\,M_{\odot}$ material. After about 3 Myr, the donor star expands again to $\sim 20\,R_{\odot}$ and  Case BC MT begins. Before this phase, the donor star has lost $\sim 0.2\,M_{\odot}$ of hydrogen envelope via a stellar wind \citep{Nugis2000} and $\sim 0.001\,M_{\odot}$ matter is accreted by the NS.  
During the Case BC MT phase, the donor star loses $\sim 0.5M_{\rm \odot}$ of its helium envelope and the NS accretes $\sim 0.001M_{\rm \odot}$ matter. Thus, the mass accreted onto the NS through Case BC MT is comparable to that through capturing a stellar wind. Here, we emphasize that the evolution of the same binaries under the assumptions of $\alpha_{\rm CE}=0.1-1.0$ always leads to merge during a CE phase. 

\subsection{Detection of ULXs in CEDPs} \label{sec:4.2}

Our simulation has shown that CEDPs appear between the stages of HMXBs and DNSs. In the Milky Way, there are over 100 HMXBs \citep{Liu2006, Fortin2023, Neumann2023}. Combined with the lifetime of $\sim 10^7\rm\,yr$, we estimate the formation rate of Galactic HMXBs to be $\gtrsim 10^{-5}\rm\,yr^{-1}$ \citep[see also a population synthesis estimation by][]{Shao2015}. According to the observations of PSR J0737-3039, \citet{Kim2015} derived the merger rate of Galactic DNSs to be $\sim 10-50\rm\,Myr^{-1}$. Since only close DNSs can evolve to merge, the formation rate of Galactic DNSs is at least a few times larger than the merger rate \citep[see e.g.,][]{Shao2018}. Considering that supernova explosions from the donor stars in HMXBs are able to disrupt a fraction of binary systems and reduce the formation of DNSs, we can roughly estimate that the event rate of CEDP ULXs is of the order $10^{-5}\rm\,yr^{-1}$ for a Milky Way-like galaxy although part of HMXBs do not evolve to enter CEDPs (see Figure \ref{14}). As these ULXs can last about $10^4-10^5\rm\,yr$, it is possible to observe them in the Local Group galaxies and beyond. 

CEDP ULXs as post-CE binaries are expected to have circle orbits. Some NS binaries with a Be star or a supergiant companion have been observed as ULXs \citep{Kaaret2017,King2023}. Similar to Galactic Be/X-ray binaries and supergiant X-ray binaries, these ULX systems are expected to have eccentric orbits. We suggest that the eccentricity measurements for the NS ULXs with wide orbits can be used as an indicator to identify CEDP systems. Since the donor stars in CEDP binaries have a bloated hydrogen envelope, in some cases they look like a (super)giant star. Our results indicate that the masses of these donor stars are distributed in the range of $\sim 2-7\,M_\odot$. For an NS ULX, the inconsistency between the inferred spectral type and the measured dynamical mass of the donor star may provide a clue to consider this ULX as a CEDP binary. Interestingly, \citet{Li2022} claimed the detection of a binary recently evolved off a CE phase and in this post-CE binary the stripped star of a hot subdwarf is transferring its matter to the accretor of a white dwarf via RLOF. This source seems to be the analogue of the CEDP binaries we proposed at the low-mass end. 

By modeling Case BB/BC MT from a naked helium star onto an NS to explain the formation of Galactic binary pulsars with an NS or a white-dwarf companion, previous works suggested the accretion efficiency of the NS is larger than the Eddington limit by a factor of $\sim 2-3$ \citep{Lazarus2014,Tauris2015}. Our simulation has shown that an NS can accrete $\sim 10^{-3}M_{\odot}$ material in a CEDP. The existence of a CEDP during binary evolution indicates that an NS has already been recycled before a possible phase of Case BB/BC MT.

\subsection{The Effect of Input Parameters on CE Evolution} \label{sec:4.3}

In this section, we evaluate how our results are impacted by the input parameters (i.e., $\dot M_{\rm high}$, $\dot M_{\rm low}$ and $\delta$) of controlling CE evolution. In our default model, we set  $\dot M_{\rm high}=1\,M_{\odot}\,\rm yr^{-1}$, $\dot M_{\rm low}=10^{\rm -5}\,M_{\odot}\,\rm yr^{-1}$ and $\delta=0.02$. Accordingly, our simulation has shown that post-CE binaries are distributed in markedly different regions (e.g., see the $M_{\rm H,env}^{\rm CE}-\log P_{\rm orb}^{\rm CE}$ diagram in Section \ref{sec:3.3}). Since modeling a large grid of binaries will cost a lot of computing resources, we only evolve some specific binaries initially containing a $10M_{\rm \odot}$ donor star. 

In the case of $\alpha_{\rm CE}=3.0$, we select two binaries with $\log (P^{\rm i}_{\rm orb}/\rm d)=2.75$ or $\log (P^{\rm i}_{\rm orb}/\rm d)=2.76$. In our default model, the former system lasts $\sim 1000$ yr during CE evolution while the latter system only lasts $\sim 5$~yr. At the termination of CE evolution, the former system has $M_{\rm He,core}^{\rm CE}=2.41\,M_{\rm \odot}$, $M_{\rm H,env}^{\rm CE}=0.45\,M_\odot$ and $P_{\rm orb}^{\rm CE}=0.16$ days, while the latter system has $M_{\rm He,core}^{\rm CE}=2.41\,M_{\rm \odot}$, $M_{\rm H,env}^{\rm CE}=2.05\,M_\odot$ and $P_{\rm orb}^{\rm CE}=19.5$ days. In the case of $\alpha_{\rm CE}=1.0$, we select another two binaries with $\log (P^{\rm i}_{\rm orb}/\rm d)=3.07$ or $\log (P^{\rm i}_{\rm orb}/\rm d)=3.08$. Since the CE phases in the former and the latter binaries respectively last $\sim 214$ yr and $\sim 5$ yr, the evolutionary consequences of two post-CE systems have a significant difference ($M_{\rm He,core}^{\rm CE}=3.63\,M_{\rm \odot}$, $M_{\rm H,env}^{\rm CE}=0.15\,M_\odot$ and $P_{\rm orb}^{\rm CE}=8.37$ days for the former or $M_{\rm He,core}^{\rm CE}=3.63\,M_{\rm \odot}$, $M_{\rm H,env}^{\rm CE}= 0.22\,M_\odot$ and $P_{\rm orb}^{\rm CE}=36.0$ days for the latter). Table \ref{tab:t1} shows the calculated results with different input parameters from the default values.
Overall, varying the input parameters can hardly affect the values of $E_{\rm bind}$ but significantly change the durations of CE evolution. Varying $\dot M_{\rm high}$ can affect the parameters of the binaries at the moment when CE evolution is triggered, while varying $\dot M_{\rm low}$ and $\delta$ can affect the parameters of post-CE binaries at the termination of CE evolution.
We see that adopting different input parameters is able to change the evolutionary fates of specific binary systems.

Next, we check the effect of the input parameters on the lower boundaries between CE survivors and CE mergers in the $M^{\rm i}_{\rm d}-\log P^{\rm i}_{\rm orb}$ diagram (see Figure \ref{1}). Also, we adopt different input parameters as presented in Table \ref{tab:t1}. We set $\alpha_{\rm CE}=1.0$ and evolve the binaries with $\log (P^{\rm i}_{\rm orb}/\rm d)=3.01-3.10$ in an interval of 0.01. In the default model, the binaries with $\log (P^{\rm i}_{\rm orb}/\rm d)\geq 3.06$ can survive CE evolution. When using different input parameters, this lower boundary of orbital period does not change in most cases. Only when $\delta=0.001$, the minimum initial orbital period shifts to $\log (P^{\rm i}_{\rm orb}/\rm d)= 3.04$.

\section{Conclusion} \label{sec:conclusion}

We have used grids of MESA simulations with updated methods for MT and CE evolution to present the evolutionary consequences from HMXBs to DNSs. Our models are computed beginning from the binary systems with a zero-age main sequence star and a $1.4M_{\rm \odot}$ NS. The initial orbital periods span a range of $2.5\leq\log (P_{\rm orb}^{\rm i}/\rm d)\leq 3.5$ in steps of 0.01, and the initial donor masses cover a range of $M_{\rm d}^{\rm i}=8-20\,M_{\odot}$ with an interval of $1\,M_{\odot}$. In our calculations, we adopt four different CE ejection efficiencies of $\alpha_{\rm CE} = 3.0$, 1.0, 0.3, and 0.1. In the following, we present our main results.

There is a tendency that the CE survivors in the $M^{\rm i}_{\rm d}-\log P^{\rm i}_{\rm orb}$ diagram occupy smaller parameter spaces with decreasing $\alpha_{\rm CE}$. Since the donor stars in post-Case B systems have $|E_{\rm bind}|$ many times larger than those in post-Case C systems, the former binaries are more likely to merge during CE evolution while the latter binaries evolve to be CE survivors.

Our calculations indicate that the post-CE binaries with relatively wide orbits can evolve to enter CEDPs and the binaries with relatively narrow orbits may avoid to enter CEDPs. In some cases, the inspiral phases of CE evolution only last a few years which allows the donor stars to remain the hydrogen envelopes of $M_{\rm H,env}^{\rm CE}\sim 1.0-4.0\,M_{\odot}$. In most cases, the inspiral phases can last about $10^3$ yr which leads to the formation of the post-CE binaries with the donor stars of $M_{\rm H,env}^{\rm CE}\lesssim 0.4-1.0\,M_{\odot}$. We notice that the input parameters of controlling CE evolution can be responsible for the hydrogen-envelope masses remained during CE phases.

Almost all donor stars in post-Case B binaries have $M_{\rm H,env}^{\rm f}=0$ due to the mass transfer in CEDPs and the mass loss via stellar winds. The donor stars in some post-Case C binaries also have $M_{\rm H,env}^{\rm f}=0$, which mainly caused by the mass loss due to the mass transfer in CEDPs. Post-Case C binaries emerge in the region where $M_{\rm H,env}^{\rm f}\sim 0.02-0.2M_{\rm \odot}$ and $P_{\rm orb}^{\rm f}\sim 0.4-600$~days.


Based on our calculations, all CE survivors are expected to have a helium star of $M_{\rm He,core}^{\rm CE}\lesssim 7\,M_{\odot}$. Since the mass of the donor star in Cygnus X-3 was suggested by \citet{Zdziarski2013} to be between $7.2\,M_{\odot}$ and $17.5\,M_{\odot}$ \citep[see also][]{Antokhin2022}, we propose that the compact object in this source is more likely to be a low-mass black hole rather than an NS.

The donor stars in some post-Case B binaries with close orbits may finally explode as ultra-stripped supernovae as they have $M_{\rm He,env}^{\rm f}\lesssim 0.2\,M_{\odot}$. The processes of both a stellar wind and RLOF MT have a significant contribution to the extreme stripping of the donor stars. Additionally, the donor stars have $M_{\rm He,env}^{\rm f}\sim 0.2-1.0\,M_{\odot}$ in most post-Case B binaries and $M_{\rm He,env}^{\rm f}\sim 0.3-1.7\,M_{\odot}$ in post-Case C binaries.

The NSs in post-Case B binaries increase masses mainly through the process of RLOF MT in CEDPs and Case BB/BC phases and the process of capturing a stellar wind \citep{Nugis2000}. The NSs in post-Case C binaries increase masses mainly through the process of capturing a stellar wind \citep{deJager1988} before the onset of RLOF MT and CE evolution. The minimum $P_{\rm orb}^{\rm f}$ of CE survivors are $\sim 0.04-2.1$ days when varying $\alpha_{\rm CE}$ from 3.0 to 0.1. Connected to the observations of close DNS systems \citep{Tauris2017}, the situations of $\alpha_{\rm CE}=3.0$ and $\alpha_{\rm CE}=0.3$ seem to be favored.

\begin{acknowledgments}
We thank the referee for helpful suggestions that improved our manuscript. YN thanks Shi-Jie Gao for his help of installation and use of MESA. This work was supported by the National Key Research and Development Program of China (grant Nos 2021YFA0718500 and 2023YFA1607902), the Natural Science Foundation of China (Nos 12041301, 12121003, and 12373034), the Strategic Priority Research Program of the Chinese Academy of Sciences (Grant No. XDB0550300). 
All input files and associated data products to reproduce our results are available for download from Zenodo at \dataset[doi:10.5281/zenodo.14184177]{https://doi.org/10.5281/zenodo.14184177}.
\end{acknowledgments}

\appendix

\section{Another evolutionary track to form a DNS system}\label{AA}

\begin{figure}[htbp]
    \includegraphics[width=0.6\textwidth]{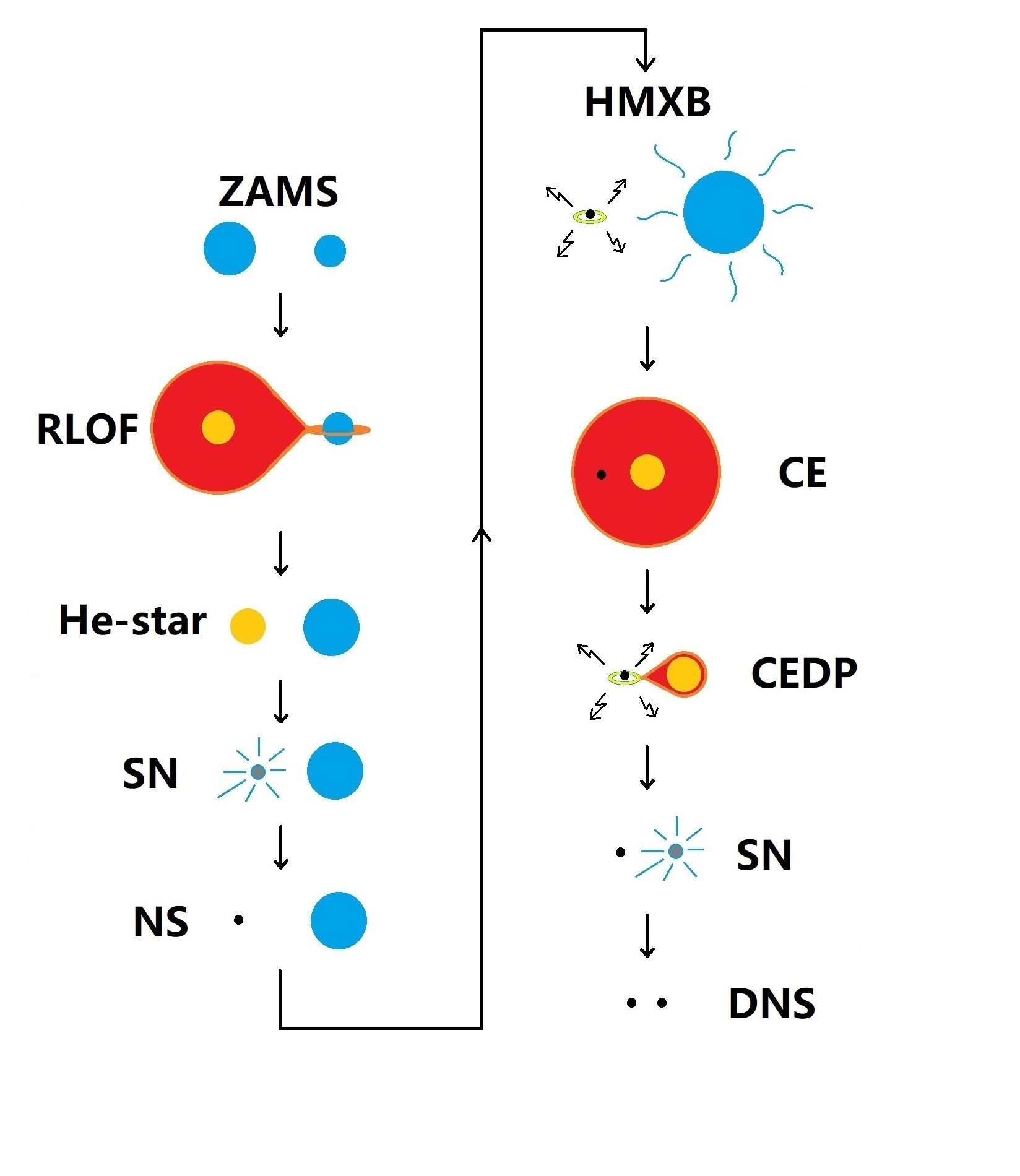}
    \centering
    \caption{Similar to Figure \ref{13}, but the post-CE binary evolves to directly form a DNS from a CEDP.}
    \label{12}
\end{figure}

Figure \ref{12} shows the formation of a DNS system originally evolved from the binary with two OB-type stars. After CE evolution, the binary evolves into a CEDP and then forms a DNS. This picture depicts a typical case for the evolution of an HMXB experienced Case C RLOF MT.

\section{Evolution of a binary without experiencing a CEDP}\label{AB}

\begin{figure*}[htbp]
    \includegraphics[width=0.9\textwidth]{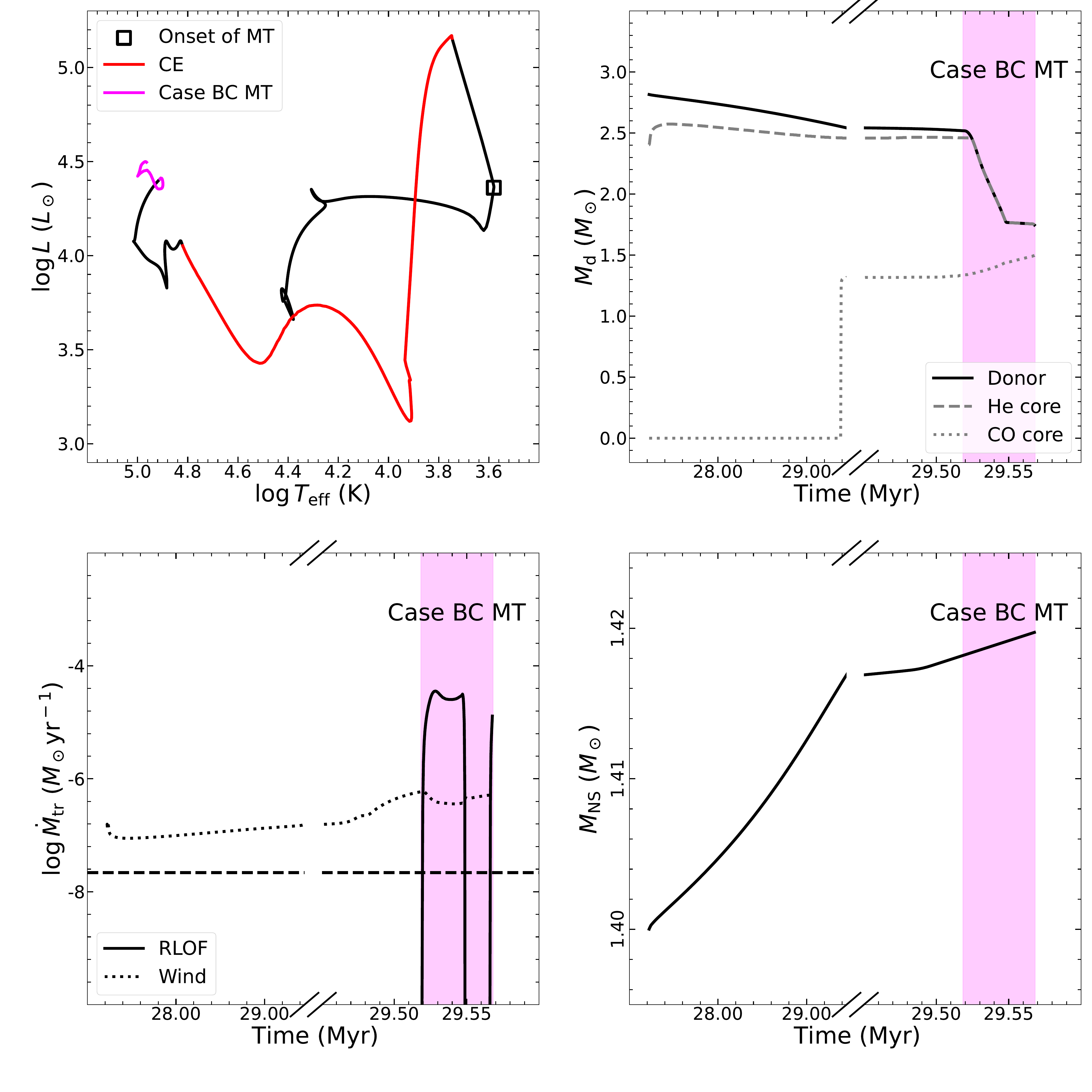}
    \centering
    \caption{Similar to Figure \ref{11}, but evolving the binary with $M_{\rm d}^{\rm i} = 10\,M_\odot$. In the lower-left panel, the dotted curve denotes the mass loss rate due to a stellar wind \citep{Nugis2000}.}
    \label{14}
\end{figure*}

In Figure \ref{14}, we show the evolutionary tracks of the binary initially containing a $1.4\,M_\odot$ NS and a $10\,M_\odot$ donor in a 537-day orbit. During CE evolution, we adopt $\alpha_{\rm CE}=3.0$. This system does not evolve into a CEDP. After CE evolution, a stellar wind directly leads the donor star to lose $\sim 0.3\,M_{\odot}$ of envelope matter in $\sim 2.3$~Myr. At the time of $\sim 29.52$~Myr, Case BC MT phase begins. During this phase,  $\sim 0.8\,M_{\odot}$ of helium envelope is stripped away and $\sim 0.002\,M_{\odot}$ matter is accreted by the NS.

\bibliography{reference}{}
\bibliographystyle{aasjournal}



\end{document}